\documentclass[11pt,a4paper]{article}
\usepackage[
  a4paper,
  left=1cm,
  right=1cm,
  top=2.5cm,
  bottom=2.5cm
]{geometry}

\usepackage[numbers,sort&compress]{natbib}

\usepackage{graphicx}
\usepackage{url}
\usepackage{hyperref}
\usepackage{amsmath, amssymb, bm}
\usepackage{subcaption}
\usepackage{multirow}
\usepackage{fancyhdr}
\usepackage{appendix}

\newcommand{\keywords}[1]{\par\noindent\textbf{Keywords: }#1}

\title{Silencing Newtonian noise using fusion sensor arrays}

\author{%
\parbox{\textwidth}{%
  \raggedright
  Paul Ophardt$^{1}$,
  Francesca Badaracco$^{2}$,
  Katharina\mbox{-}Sophie Isleif$^{1}$\\[0.5em]
  \normalsize $^{1}$Helmut Schmidt University, Institute for Mechanical Engineering and Civil Engineering, 22043 Hamburg, Germany\\
  \normalsize $^{2}$Università degli Studi di Genova, Dipartimento di Fisica (DIFI), 16146 Genova, Italy\\[0.5em]
  \normalsize \texttt{ophardtp@hsu-hh.de}
}}

\date{} % empty = no date printed

\begin{document}

\maketitle

\begin{abstract}
Newtonian noise (NN) from seismic density fluctuations is expected to limit the low-frequency sensitivity of third-generation gravitational-wave detectors, in particular the Einstein Telescope (ET). Current NN mitigation relies on seismometer arrays and Wiener filtering, while distributed acoustic sensing (DAS) offers a complementary, low-cost means of obtaining dense strain measurements.
We investigate fusion sensor arrays composed of both displacement-measuring seismometers and strain-measuring DAS-type sensors. We extend the Wiener filter formalism to mixed sensor types and introduce analytic S-wave strain correlation coefficients. Using a hybrid differential evolution and covariance matrix adaptation scheme, we validate our approach against established seismometer-only results and analyze the geometry, robustness, and performance of optimized fusion arrays.
Fusion arrays enhance P/S-wave disentanglement and achieve NN cancellation levels comparable to, and sometimes exceeding, those of seismometer-only arrays, particularly for small sensor numbers. When sensors are constrained to the ET infrastructure, we find that six seismometers complemented by fourteen strainmeters inside the ET arms can match the performance of twenty seismometers in boreholes, achieving a residual at the 10\% level, and thereby offering a cost-efficient pathway toward ET-scale NN mitigation.
\end{abstract}

\keywords{Newtonian noise, Einstein Telescope, position optimization for seismometer and strainmeter}

% ---- main text starts here ----

% Our results show that fusion sensor arrays can match or exceed the performance of traditional seismometer-only arrays. We further evaluate fusion arrays under varying P- and S-
% wave compositions, as well as for an array located in the vicinity of the ET infrastructure only. We find that, by supporting a seismometer array with a strainmeter array, which is installed inside of ET, we can reduce the number of necessary boreholes from 12 to 6, attaining the same performance. 
%\end{abstract}
%Abstract has now 164 words

\section{Introduction}

The Einstein Telescope (ET) promises new breakthroughs in gravitational-wave (GW) physics. For instance, it will enable the discovery of new GW sources, such as core-collapse supernovae  \cite{vartanyan_gravitational_2020, yuan_waveform_2024} and isolated neutron stars  \cite{sieniawska_gravitational_2021, pagliaro_searching_2025}, and further, the detector will advance multi-messenger astrophysics  \cite{koehn_impact_2024, giangrandi_numerical_2025}. The references listed above are illustrative rather than exhaustive, for a comprehensive overview of the science enabled by ET, see \cite{maggiore_science_2019}. To reach these goals, the ET's sensitivity would have to be improved by a factor of 10 compared to the current advanced detectors, and its frequency band extended down to 3 Hz \cite{punturo_einstein_2010}. The GW detector’s noise floor around 3 Hz is dominated by seismic and Newtonian noise (NN). By building the ET underground, both of these noise sources can be suppressed, since they have large surface contributions, such as, from Rayleigh waves. Nevertheless, relocating the detector to the underground on its own is not sufficient to mitigate seismic NN, as it may still limit the achievable design sensitivity \cite{harms_lower_2022}.

Therefore, in order to reach the science goals for the ET, strategies must be developed to mitigate seismic NN. Current GW detectors of the second generation already employ a NN cancellation system. The Advanced Virgo detector, for instance, has deployed an array of seismometers measuring the seismic field around the interferometer for the purpose of NN mitigation \cite{koley_design_2024}. By modeling the field with plane surface waves, for example, the NN can be calculated and removed from the data. Besides these efforts, no NN from seismic density fluctuations has been measured yet, and Newtonian noise mitigation was therefore not applied, but only tested for Virgo. While other approaches have also been evaluated for Advanced Virgo, such as recess structures \cite{harms_passive_2014, singha_newtonian-noise_2020}, mitigating NN using sensor arrays seems to be the most principled approach for the ET.

The primary challenge associated with the installation of a sensor network for the mitigation of NN is the optimal placement of the sensors. The seismometer positions for the Advanced Virgo detector were optimized using a machine learning algorithm \cite{badaracco_machine_2020}. The NN is estimated using a Wiener filter (WF), which exploits the coherent component of the seismic field across the sensor array. The model for the seismic field is calculated analytically and assumes that the seismic noise floor is composed of an isotropic distribution of sources, which are further uncorrelated and stationary \cite{harms_terrestrial_2019}. Using body instead of surface waves and a metaheuristic optimization algorithm to find the optimal positions, this seismic model was applied to optimize a sensor array mitigating NN for the ET in \cite{harms_optimization_2019, badaracco_joint_2024}. In these works, different metaheuristic optimization algorithms, such as Differential Evolution and Particle Swarm Optimizers, have been considered. Later, more sophisticated gradient-descent algorithms were studied as well \cite{schillings_fighting_2025}. Placing a sensor array underground for ET requires drilling boreholes at the designated locations. The boreholes are a major cost factor for the NN cancellation system. It was estimated in \cite{badaracco_joint_2024} that, for a cancellation of NN acceleration of a factor 10, hundreds of boreholes would be necessary, which results in an unfeasible budget.

In this work, in order to improve NN cancellation systems and reduce these costs, we investigate alternative sensor arrays. In addition to seismometers, we consider another type of sensor, namely, strainmeters. While seismometers measure ground displacement, strainmeters measure the covariant derivative of displacement, namely, strain. By using an array which is composed of different types of sensors, we are able to improve disentanglement of the information about the seismic fields, P- and S- wave content. This is particularly interesting in the WF for a small number of sensors. We call these sensor arrays \textit{fusion sensor arrays}.

One promising type of strainmeter utilizes so called distributed acoustic sensing (DAS), which is a commercial device containing a pulsed laser source which turns telecommunication glass fiber into a dense sensor network, of one sensor per one gauge length (which is usually about 10 m). This results in a dense array of sensors, covering large areas spanning up to 100 km. Although DAS sensors do not yet match seismometers in sensitivity, active research efforts \cite{miah_review_2017, zinsou_recent_2019, lu_distributed_2019, wang_recent_2020, bölt2025distributedacousticfibersensing} are driving rapid improvements, suggesting substantial future potential. DAS offers significant potential for NN cancellation, as it can complement seismometers by helping to disentangle the P- and S-wave content, provide spatially dense coverage of seismic wavefields, and monitor extended sections of the ET infrastructure, including anthropogenic sources such as machinery.

Furthermore, we investigate alternative installations for the sensor arrays, namely arrays which are located partly inside of the ET infrastructure, in contrast to arrays fully installed in separate boreholes. Hence, we restrict the solution space for the sensors to be inside of the ET arms, in contrast to optimizing for the global minimum. This is beneficial, as it can potentially reduce the costs of drilling boreholes. We evaluate different configurations for the arrays, either constraining the whole array to the ET or just the strainmeters. 

In section \ref{sec: methods}, we introduce the WF method and show how to integrate strainmeter sensors into the formalism, by calculating and interpreting their correlation coefficients. In the same section, we introduce the algorithm, which is used to find the optimal positions for the sensors, maximizing the NN cancellation capabilities. We further validate the algorithm, and investigate the solutions for fusion sensor arrays with a total of 6 sensors. After having validated our algorithm, we evaluate the capabilities of fusion sensor arrays to reach the targeted NN cancellation of a factor of 10 in \ref{sec: FA evaluation}, by analyzing larger numbers of sensors and compare the results to the performance of pure seismometer arrays. Also targeting the design sensitivity of the ET, we evaluate fusion sensor arrays partly constrained to the ET infrastructure in section \ref{sec: FA in ET}.

\section{Methods}\label{sec: methods}
\subsection{Fusion sensor arrays in the Wiener filter}\label{subsec:FusionsensarraysinWF}
In this section we explain how to integrate strainmeter sensors into the WF formalism. Using the measurements from a sensor array, we employ the WF method to cancel NN from seismic body waves from a test mass (TM) situated underground and surrounded by a cavity. For a direct comparison, we follow the configuration introduced by \textit{Badaracco et al.} \cite{harms_optimization_2019}, where NN is cancelled for a single TM placed at the center of a spherical cavern. We assume that the spherical radius is small compared to the seismic wavelength, and neglect scattering from the walls. Further, we assume that the medium around the cavity is infinite and homogeneous, with the density $\rho_0$. The analytical calculation of NN for a spherical cavern, using all of the assumptions mentioned above, is detailed in  \cite{harms_terrestrial_2019}, and the equation for the NN acceleration at the TM reads
\begin{equation}\label{eq:nn_solution}
    \delta \mathbf{a}(\mathbf{r}_0,t) = \frac{4\pi}{3}\, G\,\rho_0\, \big(2\,\boldsymbol{\xi}^{\mathrm{P}}(\mathbf{r}_0,t) - \,\boldsymbol{\xi}^{\mathrm{S}}(\mathbf{r}_0,t)\big)\,.
\end{equation}
The seismic displacement fields $\boldsymbol{\xi}^P$ and $\, \boldsymbol{\xi}^S$ describe the decompositions into compressional P-waves and shear S-waves, respectively, and the constant $G$ the gravitational constant.

In the context of seismic fields, monochromatic plane waves are employed. These waves can be characterized by the displacement field
\begin{equation}\label{seismic displacement fields}
    \boldsymbol{\xi}(\mathbf{r}, t) = \mathbf{p}\, e^{-i(\mathbf{r}\cdot\mathbf{k} - \omega t)},
\end{equation}
which is defined by the polarization vector $\mathbf{p}$, the wave frequency $\omega$, and the wavevector $\mathbf{k}$. We assume a random seismic noise floor, and further constrain it to be isotropic, stationary and unpolarized. We impose this by averaging the correlations between signals over the wave direction $\mathbf{e}_{k} = \frac{\mathbf{k}}{k}$ and the polarization $\mathbf{p}$, a formalism developed for the GW stochastic background \cite{allen_detecting_1999, flanagan_sensitivity_1993}.

In order to predict the seismic NN acceleration at the TM, we use an array of sensors placed optimally within the medium around the TM. NN mitigation by coherent noise cancellation from sensor arrays was first discussed in \cite{casciaro_off-line_2000}. The WF framework was then extended by optimizing the positions of the sensors in the array \cite{harms_optimization_2019, badaracco_joint_2024, schillings_fighting_2025,driggers_subtraction_2012,coughlin_towards_2016}. We build on these publications and adopt the theoretical framework for the WF, based on a random, isotropic body-wave seismic field, developed in \cite{harms_terrestrial_2019}.

The WF can be evaluated by calculating the residual 
\begin{equation}\label{eq:WF_residual}
    \mathcal{R} \;=\; 1 \;-\; \frac{\mathbf{C}_{\!SN}^{T}\, \mathbf{C}_{\!SS}^{-1}\, \mathbf{C}_{\!SN}}{C_{\!NN}}\,,
\end{equation}
where $\mathbf{C}_{\!SS}\in\mathbb{R}^{N\times N}$ is the cross-correlation matrix including all correlations between the sensor signals, $\mathbf{C}_{\!SN}\in\mathbb{R}^{N}$ is the cross-correlation vector including the correlations between the signals of each sensor and the NN acceleration at the TM, and $C_{\!NN}$ is the NN acceleration autocorrelation at the TM. The residual $\mathcal{R}$ represents the amplitude ratio of the residual NN to the original NN. A value of 0.1 corresponds to a factor of 10 cancellation. We include the signal-to-noise-ratio (SNR) of the sensors into the analysis by multiplying the diagonal terms of the cross-correlation matrix by $(1 + 1/\text{SNR}^2)$. 

The WF method combines the signals measured in the sensors around the TM coherently, and it is the optimal method based on the fact that the NN acceleration and seismic fields are related linearly \cite{vaseghi_advanced_2000, coughlin_wiener_2014}, as seen in equation \eqref{eq:nn_solution}. Analyzing in Fourier domain, the same holds for the seismic strain field 
\begin{equation}\label{seismic strain fields}
    \bm{\epsilon}(\mathbf{r}, t) = \partial_\mathbf{r} \boldsymbol{\xi}(\mathbf{r},t) = -i \, \mathbf{p}\otimes \mathbf{k} \, e^{-i(\mathbf{r}\cdot\mathbf{k} - \omega t)},
\end{equation}
justifying the inclusion of strainmeter sensors in the WF formalism.

By adding strainmeters to the WF formalism, the resulting fusion array automatically includes all cross-correlations, including those between strainmeters and seismometers. Appendix \ref{appendix:fusionWF} provides a concrete example of how these mixed correlations are computed. Since the two sensor types exhibit distinct noise characteristics, combining their signals requires careful treatment \cite{rading_distributed_2025, paitz_empirical_2021, shinohara_performance_2022, rossi_assessment_2022, kasahara_comparison_2018}.

The two-point correlation functions entering the WF method (e.g. the cross-correlation matrix) can be calculated as outlined in \cite{harms_terrestrial_2019}. Because the NN acceleration at the TM is directly proportional to the displacement of the seismic wave, the correlation coefficients between the sensors and the NN signal, collected in $\mathbf{C}_{\!SN}$, can be calculated using the correlation coefficients of the respective sensor and wave type multiplied by the factor $\frac{4}{3}\pi G\rho_0$. A comprehensive summary of all correlation coefficients is given in table \ref{tab:correlation_coefficients_explicit}. In this table, we newly introduce all strainmeter correlation coefficients for S-waves. Compared with the formulation in \cite{harms_terrestrial_2019} (Eqs. 207, 209, 214, 215), we refrain from normalizing the degrees of freedom. In that article, displacement coefficients were multiplied by 3 and strain coefficients by 5. Because these differing normalization factors become inconsistent when displacement and strain channels are used jointly in the Wiener filter, we do not apply them in our approach.

\begin{figure}[!htbp]
    \centering
    \includegraphics[width=0.6\linewidth]{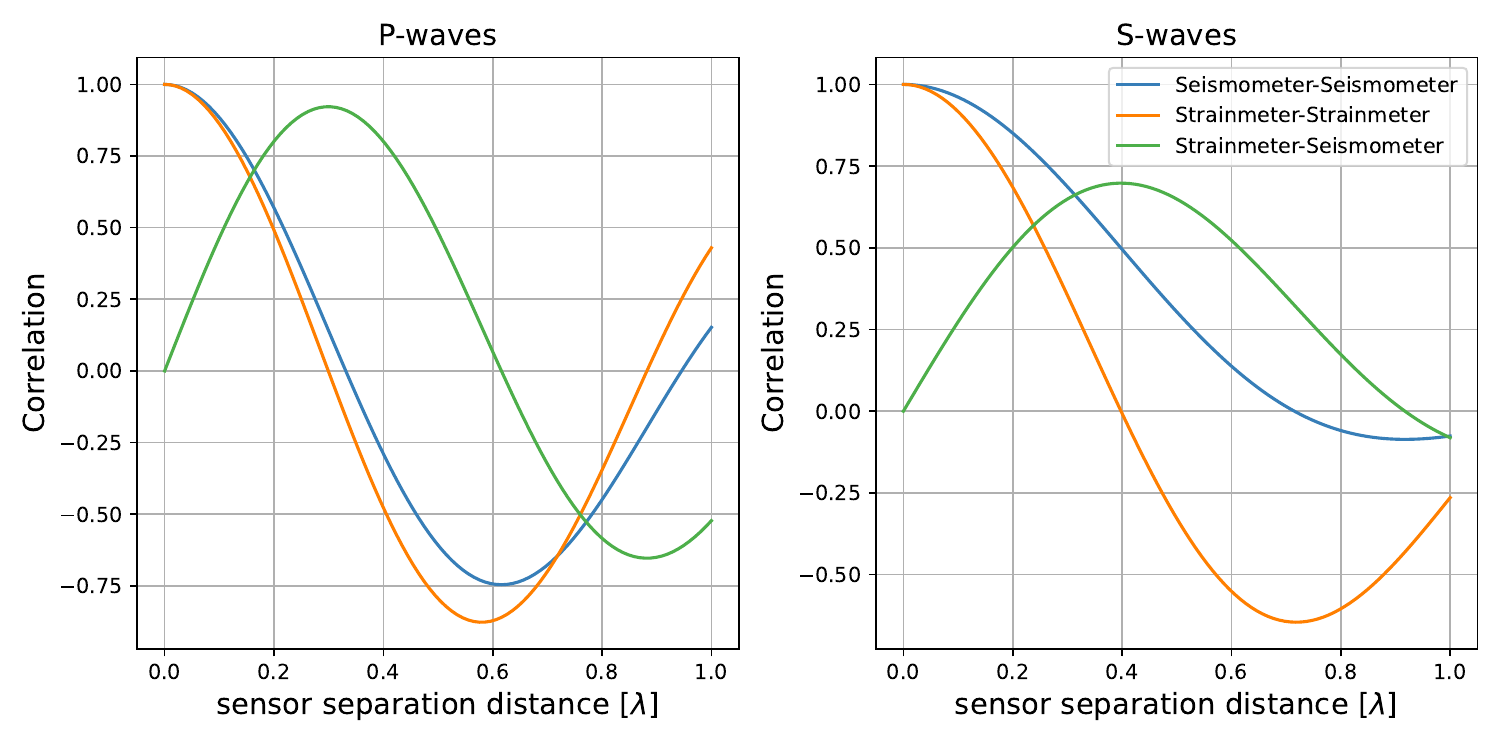}
    \caption{Comparison of the different correlation types between two sensors, including seismometer and strainmeter. The x-ordinate shows their separation distance $r$, normalized by the wavelength $\lambda$, for the seismic P- and S-wavelengths respectively. Both sensor types measure along the same direction, chosen to be parallel to their separation vector ($\mathbf{e}_1 = \mathbf{e}_2 = \mathbf{e}_{12}$). All of the correlations are normalized such that the auto-correlation between same sensor types are equal to one. Instrumental noise is neglected (infinite SNR).}
    \label{fig:Compare strain to seismometer correlations}
\end{figure}

\begin{table}[!htbp]
\renewcommand{\arraystretch}{2}
\centering
\caption{Two-point correlation coefficients between two sensors for seismic plane wave fields from \cite{harms_terrestrial_2019} (note that they are normalized differently here, adapted to a WF where different sensor types are included), newly introducing all coefficients for strainmeter measuring S-waves. The unit vector connecting the sensor positions $\mathbf{r}_1$ and $\mathbf{r}_2$ is defined as $\mathbf{e}_{12} = (\mathbf{r}_2 - \mathbf{r}_1)/|\mathbf{r}_2 - \mathbf{r}_1|$ and the separation distance as $r = |\mathbf{r}_2 - \mathbf{r}_1|$. The vectors $\mathbf{e}_1, \mathbf{e}_2$ define the measurement direction of the sensors. Spherical Bessel functions $j_n(k\cdot r)$ of degree $n$ take $k\cdot r$ as the argument, with the wavenumber $k$ of the respective P- or S-wave, and they are simplified in notation by $j_n$. For the \textit{Seismometer-Strainmeter} correlation, $\mathbf{e}_1$ defines the measurement direction of the strain sensor, and $\mathbf{e}_2$ for the seismometer.}
\vspace{0.5em}
\begin{tabular}{lll}
\hline
\textbf{Correlation Type} & \textbf{Wave Type} & \textbf{Isotropic Correlations} \\
\hline
\noalign{\vskip 4pt}

\multirow{2}{*}{\textit{Seismometer-Seismometer}} 
& P-wave & 
$\begin{aligned}
(j_0 + j_2)(\mathbf{e}_1 \cdot \mathbf{e}_2) 
- 3j_2(\mathbf{e}_1 \cdot \mathbf{e}_{12})(\mathbf{e}_2 \cdot \mathbf{e}_{12})
\end{aligned}$ \\

& S-wave & 
$\begin{aligned}
(j_0 - \tfrac{1}{2}j_2)(\mathbf{e}_1 \cdot \mathbf{e}_2) 
+ \tfrac{3}{2}j_2(\mathbf{e}_1 \cdot \mathbf{e}_{12})(\mathbf{e}_2 \cdot \mathbf{e}_{12})
\end{aligned}$ \\[6pt]

\hline

\multirow{2}{*}{\textit{Strainmeter-Strainmeter}} 
& P-wave & 
\rule{0pt}{7.5ex}
$\begin{aligned}
&\tfrac{1}{105}(7j_0 + 10j_2 + 3j_4)(1 + 2(\mathbf{e}_1 \cdot \mathbf{e}_2)^2) 
+ j_4 (\mathbf{e}_1 \cdot \mathbf{e}_{12})^2(\mathbf{e}_2 \cdot \mathbf{e}_{12})^2 \\ 
&\quad - \tfrac{1}{7}(j_2 + j_4)\Big[(\mathbf{e}_1 \cdot \mathbf{e}_{12})^2 
+ (\mathbf{e}_2 \cdot \mathbf{e}_{12})^2 
+ 4(\mathbf{e}_1 \cdot \mathbf{e}_2)(\mathbf{e}_1 \cdot \mathbf{e}_{12})(\mathbf{e}_2 \cdot \mathbf{e}_{12})\Big]
\end{aligned}$ \\

& S-wave & 
\rule{0pt}{6.5ex}
$\begin{aligned}
&\tfrac{1}{630}(14j_0 + 5j_2 - 9j_4)(1 + 2(\mathbf{e}_1 \cdot \mathbf{e}_2)^2) 
- \tfrac{1}{2}j_4(\mathbf{e}_1 \cdot \mathbf{e}_{12})^2(\mathbf{e}_2 \cdot \mathbf{e}_{12})^2 \\
&\quad - \tfrac{1}{84}(-j_2 + 6j_4)\Big[(\mathbf{e}_1 \cdot \mathbf{e}_{12})^2 
+ (\mathbf{e}_2 \cdot \mathbf{e}_{12})^2 
+ 4(\mathbf{e}_1 \cdot \mathbf{e}_2)(\mathbf{e}_1 \cdot \mathbf{e}_{12})(\mathbf{e}_2 \cdot \mathbf{e}_{12})\Big]
\end{aligned}$ \\[16pt]

\hline
\noalign{\vskip 4pt}

\multirow{2}{*}{\textit{Seismometer-Strainmeter}} 
& P-wave & 
$\begin{aligned}
\tfrac{1}{5}(j_1 + j_3)\Big[(\mathbf{e}_2 \cdot \mathbf{e}_{12}) 
+ 2(\mathbf{e}_1 \cdot \mathbf{e}_{12})(\mathbf{e}_1 \cdot \mathbf{e}_2)\Big] 
- j_3(\mathbf{e}_1 \cdot \mathbf{e}_{12})^2(\mathbf{e}_2 \cdot \mathbf{e}_{12})
\end{aligned}$ \\

& S-wave & 
$\begin{aligned}
\tfrac{1}{30}(2j_1 - 3j_3)\Big[(\mathbf{e}_2 \cdot \mathbf{e}_{12}) 
+ 2(\mathbf{e}_1 \cdot \mathbf{e}_{12})(\mathbf{e}_1 \cdot \mathbf{e}_2)\Big] 
+ \tfrac{1}{2}j_3(\mathbf{e}_1 \cdot \mathbf{e}_{12})^2(\mathbf{e}_2 \cdot \mathbf{e}_{12})
\end{aligned}$ \\[6pt]

\hline
\end{tabular}
\label{tab:correlation_coefficients_explicit}
\end{table}
\noindent
Key differences between the seismometer and strainmeter correlation coefficients shown in table \ref{tab:correlation_coefficients_explicit} are as follows: By analyzing the displacement and strain fields from equations \eqref{seismic displacement fields} and \eqref{seismic strain fields}, we find that the fields differ by their amplitude and phase. They are phase shifted by $90^\circ$ from the factor $i$, and the strain field $\bm{\epsilon}$ is a rank two tensor, while the displacement field $\boldsymbol{\xi}$ is only a rank one tensor. 
Figure \ref{fig:Compare strain to seismometer correlations}, which illustrates the different coefficients from table \ref{tab:correlation_coefficients_explicit}, shows that the correlation between the seismometer and strainmeter is maximal at a sensor separating distance of around $0.3\lambda_P$ and $0.4\lambda_S$ respectively, in contrast to the auto-correlation between the sensors, which is maximal at zero distance. This comes from the $90^\circ$ phase shift between the seismic displacement and strain fields. 

\begin{figure}[!htbp]
    \centering
    \includegraphics[width=0.7\linewidth]{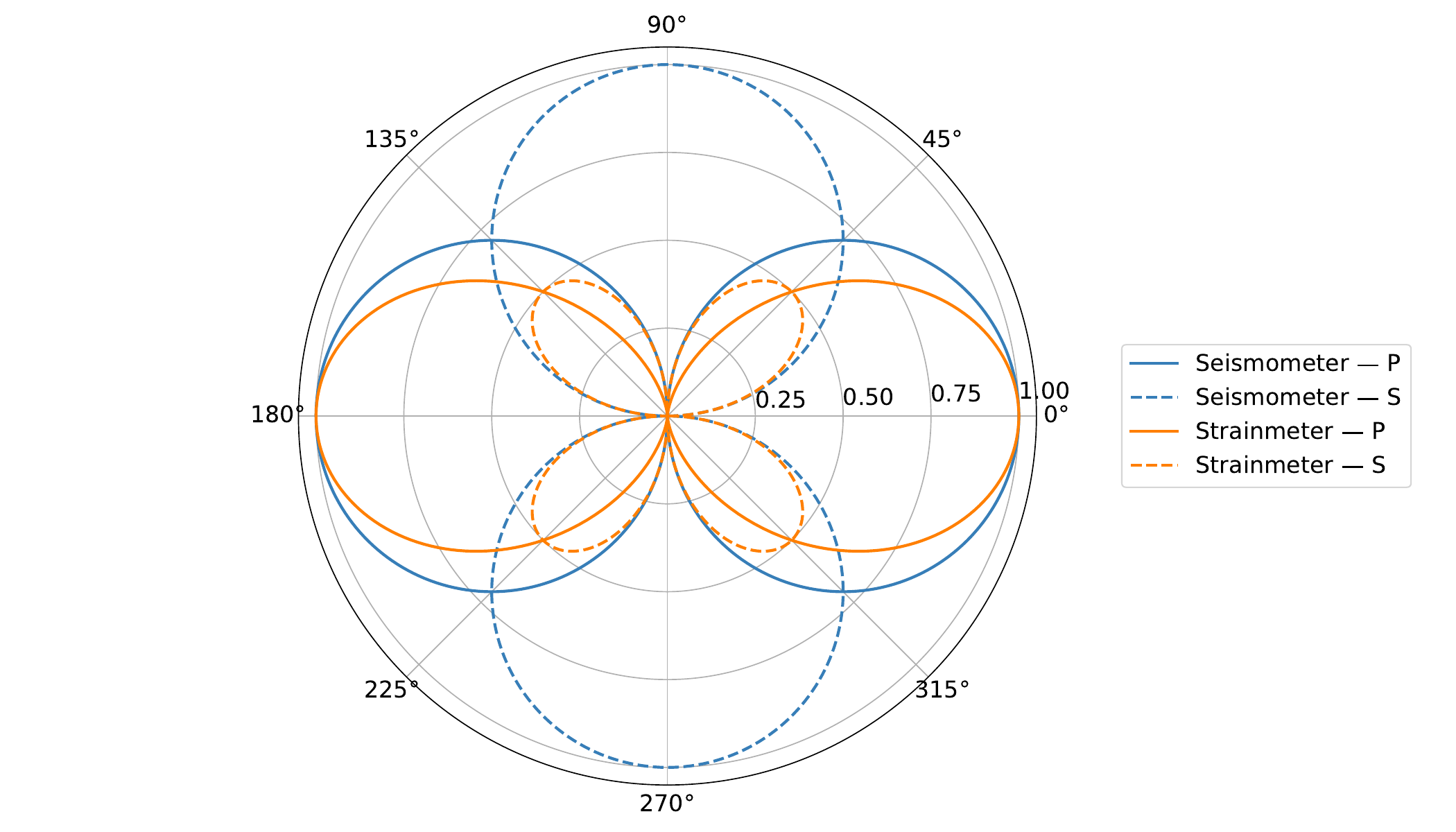}
    \caption{This figure shows the angular sensitivity of the seismometer and strainmeter sensors, for P- and S-waves, respectively. The angular sensitivity of strainmeter sensors is given by $(\mathbf{e}_1 \cdot \mathbf{e_\xi})^2$, for seismometer by $\mathbf{e}_1 \cdot \mathbf{e_\xi}$. The angles are defined as the angle between the wave-propagation direction and the sensor’s measurement direction.}
    \label{fig:Orbitals}
\end{figure}

\noindent
Measuring the strain field $\mathbf{e}_1 \cdot \boldsymbol{\epsilon}\cdot\mathbf{e}_1$, the measurement direction $\mathbf{e}_1$ enters twice as the field is a rank two tensor. For P-waves, the directionally dependent part of the amplitude of the measured strain field is given by $(\mathbf{e}_1 \cdot \mathbf{e_k})^2 = \text{cos}^2(\theta)$ and for S-waves (whose amplitude $\boldsymbol{\xi}$ is perpendicular to $\mathbf{k}$) by $(\mathbf{e}_1 \cdot \mathbf{e_\xi})(\mathbf{e}_1 \cdot \mathbf{e_k}) = \text{cos}(\theta)\,\text{sin}(\theta)$. This results in multifaceted directional sensitivity patterns, compared to the seismometers which have a directionality pattern of $\text{cos}(\theta)$ for P- and $\text{sin}(\theta)$ for S-waves. We illustrate the directional sensitivity in figure \ref{fig:Orbitals}, plotting the directional dependent part of the amplitude. Note that strainmeters are most sensitive to S-waves incident at angles of $\pm 45^\circ$, while being insensitive to waves which propagate parallel or perpendicular to the strainmeters' axis of orientation.

\newpage
\subsection{Optimization algorithm}\label{subsec: optmization alg}

The main challenge in using sensor arrays and Wiener filtering for seismically induced, NN mitigation lies in finding the optimal placement of the sensors, which minimizes the residual $\mathcal{R}$ defined in equation \eqref{eq:WF_residual}. Mathematically, this task can be formulated as an optimization problem, where, for the WF residual function $\mathcal{R}(\{\mathbf{x}\}_i)$ and the set $(\mathbb{R}^3)^N$, we seek to find the sensor positions $\{{\mathbf{x}_0\}}_i$, for which 
\begin{equation} \label{eq: minimization problem}
    \mathcal{R}(\{{\mathbf{x}_0\}_i) \leq \mathcal{R}(\{{\mathbf{x}\}}_i)}, \space \forall \{{\mathbf{x}\}}_i \in (\mathbb{R}^3)^N.
\end{equation}
For NN from Rayleigh waves, optimizing seismometer arrays was first addressed in \cite{driggers_subtraction_2012, coughlin_towards_2016}, targeting second-generation GW detectors, and later for NN from seismic body waves in \cite{harms_optimization_2019, schillings_fighting_2025, badaracco_joint_2024, jose_optimization_2021}, with a focus on the third-generation ET. Here, we extend the analysis to arrays composed of both seismometers and strainmeters.

Finding the optimal sensor placement is challenging because the problem exhibits a high number of degrees of freedom, scaling linearly with the number of sensors ($\text{dim}((\mathbb{R}^3)^N) = 3N$), and because the WF residual function defines a rough optimization landscape with many attraction basins. To address this, different optimization algorithms have been investigated, including metaheuristic approaches such as Differential Evolution (DE) \cite{storn1997differential} and Particle Swarm Optimization (PSO) \cite{kennedy1995pso}, as well as gradient-based methods such as Adam \cite{kingma2015adam}. In addition, Basin Hopping \cite{wales1997basinhopping}, which combines gradient descent with Monte Carlo steps, has been applied. A comprehensive comparison of algorithmic performance is provided in \cite{schillings_fighting_2025}. In their work, the best-performing approach, in terms of computational efficiency and robustness in locating the global minimum, was found to be a hybrid method combining PSO with Adam. More generally, \cite{schillings_fighting_2025} showed that it is advantageous to combine a metaheuristic algorithm, which can efficiently explore large regions of the solution space, with a gradient-based method, which refines the solution once the minimum is approached.

In this work, we test a new optimization scheme that combines the Differential Evolution algorithm with the Covariance Matrix Adaptation Evolution Strategy (CMA-ES) algorithm \cite{hansen2016cma}, which was easier to implement compared to gradient based methods while faster than using DE solely. While CMA-ES is a metaheuristic algorithm, it is designed to efficiently converge to a minimum close to its initial configuration. As an evolutionary algorithm, it operates on a “population” of sensor configurations and evolves to the next generation by shifting the population mean toward the configurations with the lowest residual. Its distinctive feature is the adaptive adjustment of the population spread, i.e., the covariance matrix of the shifted distribution. In addition, we employ an early-stopping criterion for the DE algorithm. The optimization algorithm is computationally efficient and straightforward to implement, but it requires careful fine-tuning of the bounds in which the DE optimization is performed. Further details on the algorithmic hyperparameters, along with a performance evaluation, are provided in the appendix \ref{appendix: CMA-ES eval}.

\section{Results}\label{sec: Results}

In this section, we present the findings of our work, regarding fusion sensor arrays which are optimized to achieve the best possible NN cancellation. In section \ref{sec: Validation}, we validate our newly developed algorithm by comparing it to previous results, obtained for seismometer arrays, and explore the geometries of the solutions as well as their stability to small perturbations in position. In section \ref{sec: FA evaluation}, we evaluate the capabilities of fusion sensor arrays by comparing it to the traditional seismometer arrays. In section \ref{sec: FA in ET}, we analyze fusion sensor arrays which are partly constrained to the ET infrastructure, and compare them to the solutions which are not constrained, hence would be placed in separate boreholes exclusively.

\subsection{Validation}\label{sec: Validation}

\begin{figure}[!htbp]
    \centering
    \includegraphics[width=0.6\linewidth]{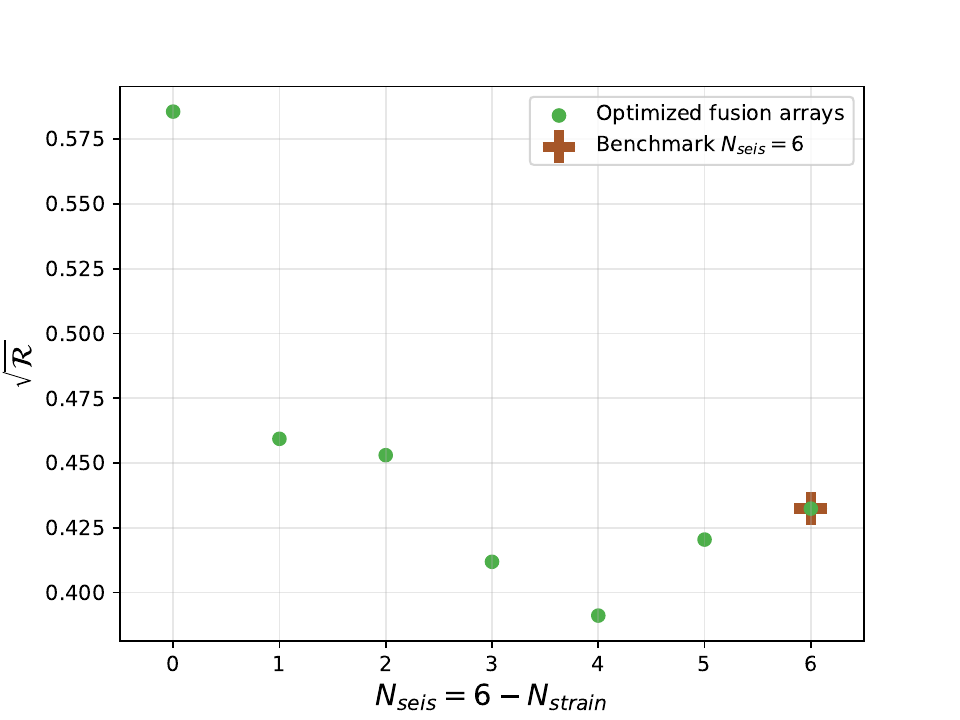}
    \caption{This figure shows the square root of the optimal WF residual, $\sqrt{\mathcal{R}}$, for different compositions of strainmeters ($N_{\mathrm{\text{strain}}}$) and seismometers ($N_{\mathrm{\text{seis}}}$) in a 6-sensor array. We assume a mixing value of $p = \frac{1}{3}$ and use an SNR of 15 for all sensors. All sensors are one axis sensors, which measure the same direction as the TMs' measurement direction. The benchmark value corresponds to $\sqrt{\mathcal{R}}$ reported in  \cite{harms_optimization_2019}.}
    \label{fig:compositionfusionarrayN6}
\end{figure}

We validate our optimization algorithm, described in \ref{subsec: optmization alg}, by comparing its performance to results previously obtained for NN mitigation from body waves using seismometer arrays in  \cite{harms_optimization_2019}. In that study, a benchmark case was defined by optimizing for 6 single-axis seismometers with an SNR of 15. For the seismic noise floor, the mixing ratio $p = \frac{1}{3}$ and the S- and P-wavelength ratio
$\lambda_s = 0.67 \cdot \lambda_p$ were used. The TM, as well as the sensors,  measure in the same direction, which is chosen to be parallel to the $x$-axis. We show that our optimizer finds this same minimum in figure \ref{fig:compositionfusionarrayN6}.

Beyond validation, in figure \ref{fig:compositionfusionarrayN6}, we also investigate the performance of the WF method using a total of 6 sensors, with different compositions of strainmeters and seismometers. Therefore, the $x$-ordinate shows the number of seismometers, which are composed of strainmeters such that the sum over all sensors is equal 6. For the strainmeters, we assume the same signal-to-noise-ratio as for the seismometers, namely $\text{SNR} = 15$. While we did so for simplicity, it is unclear whether strainmeters will reach a sensitivity comparable to those of seismometers in the future. However, as mentioned before, fiber-optical technologies are evolving and new approaches may lead to significant improvements in the near future as we move toward the construction of ET. In figure \ref{fig:compositionfusionarrayN6}, for $N_{\text{seis}} = 0$, which represents an array consisting only of strainmeters, performance is worse compared to the other configurations. The reason for this is that strainmeter sensors are less sensitive to NN than seismometers. However, fusion arrays, which have the advantage of better disentangling the P- and S-wave content compared to monotype arrays, generally perform similarly to the seismometer-only array and can even outperform it in some configurations (e.g., $N_{\text{seis}} = 3$ by up to $3\%$).

In the WF framework, canceling NN from body waves as described in section \ref{subsec:FusionsensarraysinWF}, seismometer arrays are limited only by the SNR and the mixing value between P- and S-waves. If there are only P-waves ($p = 1$) or only S-waves ($p = 0$), the array can cancel NN almost perfectly (for reasonable SNR). The reason why fusion sensor networks perform well is that the strainmeters can help seismometers to disentangle the P- and S-wave content \cite{harms_terrestrial_2019}. As explained in section \ref{sec: methods}, strainmeter sensors react differently to P- and S-waves, and therefore their measurements complement the information obtained from the seismometer measurements. In the WF, for a sensor array with a small number of sensors, the main limitation stems from a lack of information about the seismic field, therefore fusion sensor arrays are particularly useful there.

\begin{figure}[!htbp]
    \centering
    \begin{subfigure}[b]{0.22\textwidth}
        \includegraphics[width=\textwidth]{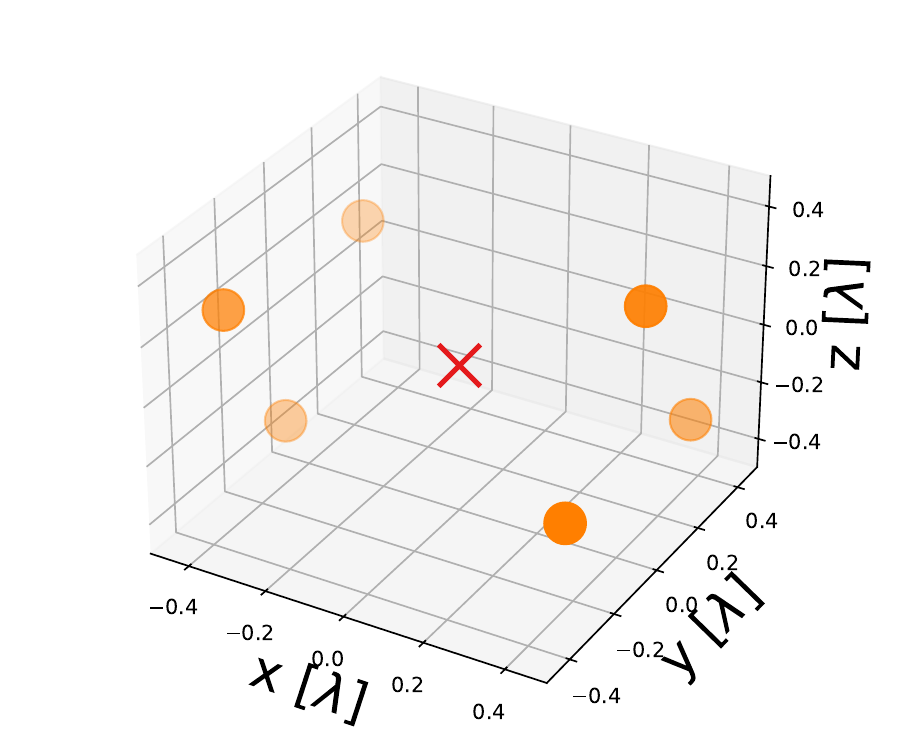}
        \caption{}
    \end{subfigure}
    \begin{subfigure}[b]{0.22\textwidth}
        \includegraphics[width=\textwidth]{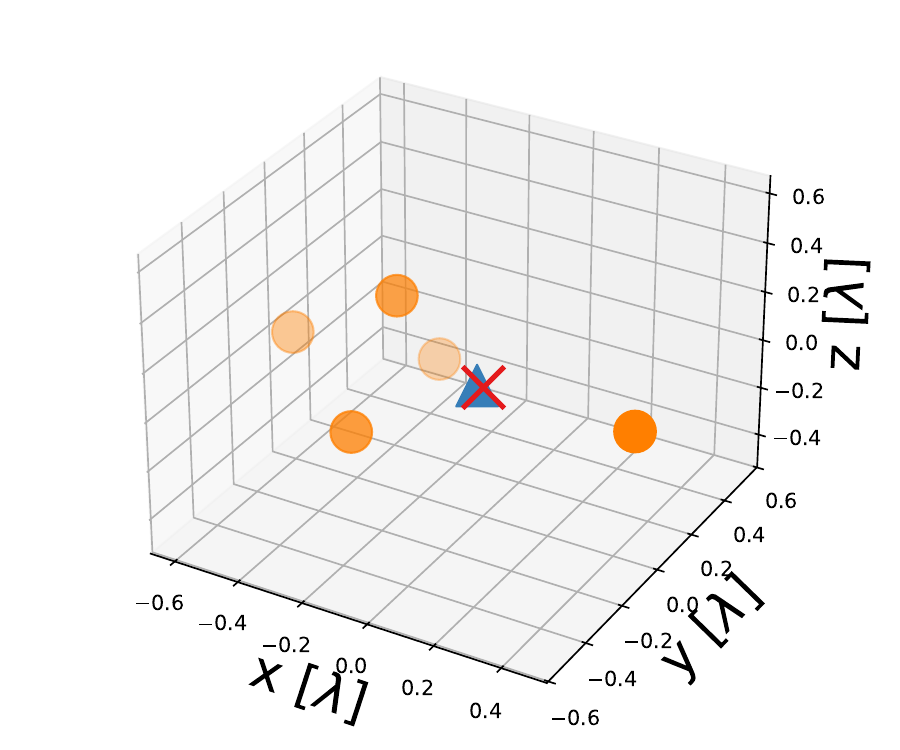}
        \caption{}
    \end{subfigure}
    \begin{subfigure}[b]{0.23\textwidth}
        \includegraphics[width=\textwidth]{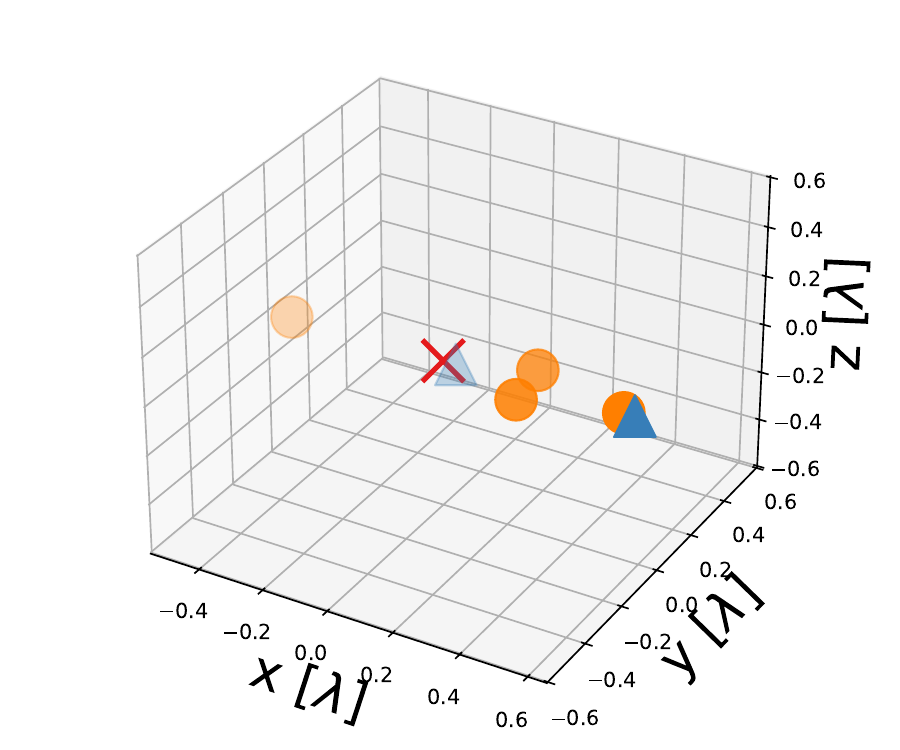}
        \caption{}
    \end{subfigure}
    \begin{subfigure}[b]{0.23\textwidth}
        \includegraphics[width=\textwidth]{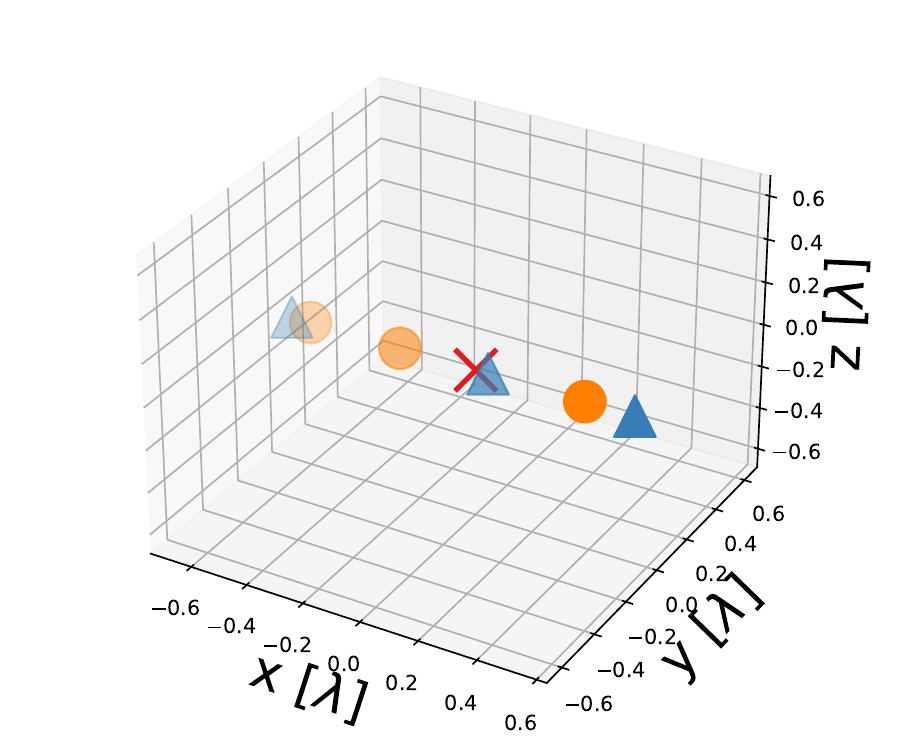}
        \caption{}
    \end{subfigure}
    \begin{subfigure}[b]{0.23\textwidth}
        \includegraphics[width=\textwidth]{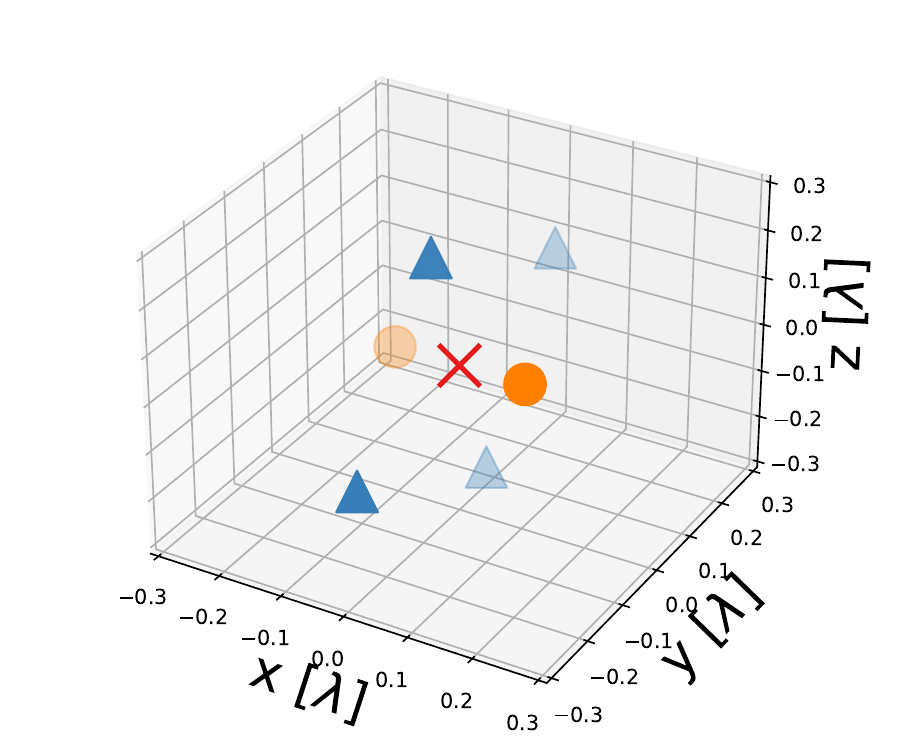}
        \caption{}
    \end{subfigure}
    \begin{subfigure}[b]{0.23\textwidth}
        \includegraphics[width=\textwidth]{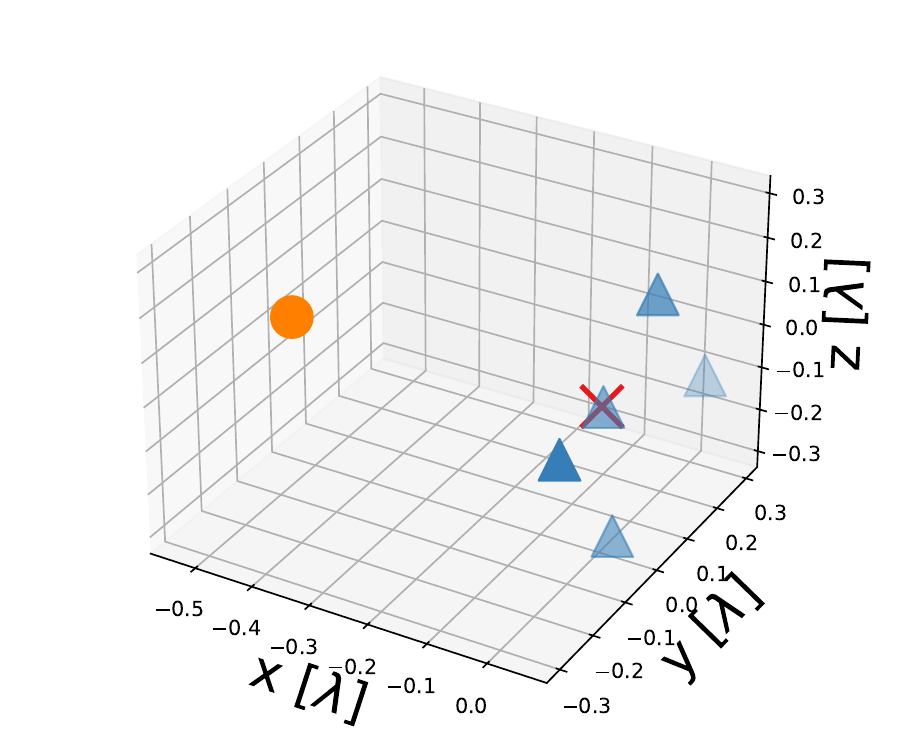}
        \caption{}
    \end{subfigure}
    \begin{subfigure}[b]{0.23\textwidth}
        \includegraphics[width=\textwidth]{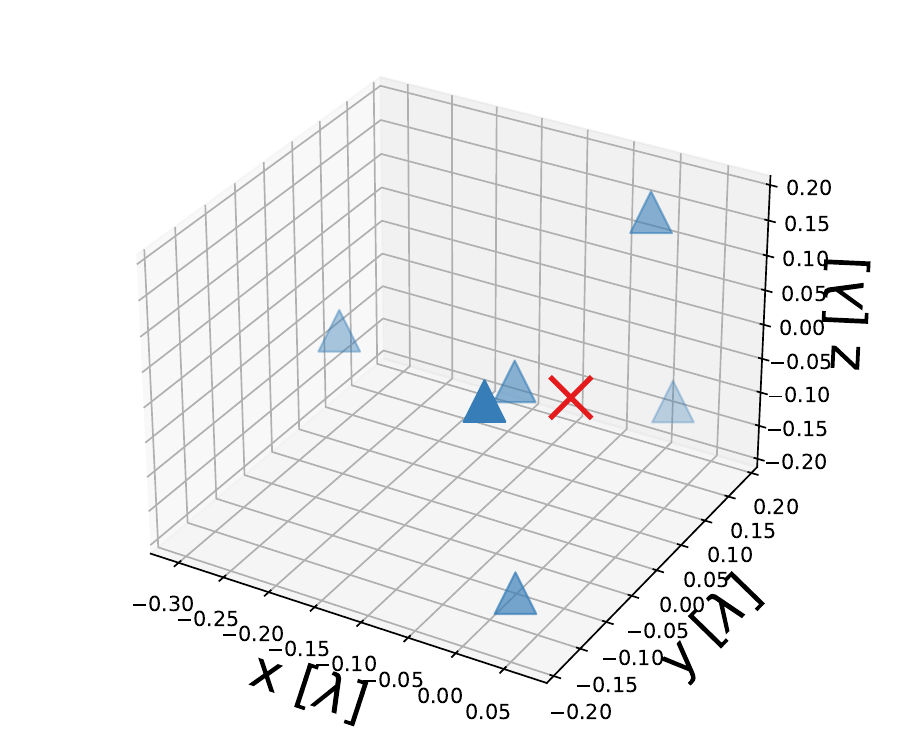}
        \caption{}
    \end{subfigure}
    \begin{subfigure}[b]{0.23\textwidth}
        \includegraphics[width=\textwidth]{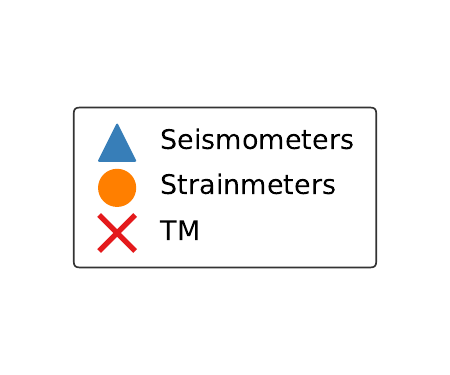}
        \caption*{}
    \end{subfigure}
    \caption{Sensor array geometries, which are optimized the to the WF filter residual. The TM measures along the $-x$ direction, and all sensors are aligned to measure in the same direction.}
    \label{fig:arraygeometries}
\end{figure}

We examine the solutions to the minimization problem defined in equation \eqref{eq: minimization problem} in greater detail by analyzing the corresponding sensor array geometries of the solutions, which we show in figure \ref{fig:arraygeometries}. The geometry of the array composed solely of seismometers in figure \ref{fig:arraygeometries}g) matches the geometry of the benchmark solution reported in  \cite{harms_optimization_2019} up to rotation around the $x$-axis, which, however, is a symmetry axis of the problem. The pure strainmeter array shown in figure \ref{fig:arraygeometries}a) is placed farther from the TM, consistent with the strain to displacement correlation coefficient, which peaks at a distance of approximately $0.3 \lambda_p$ to $0.4 \lambda_s$, see figure \ref{fig:Compare strain to seismometer correlations}. The fusion array geometries reveal a rich variety of possible configurations and indicate that co-locating sensors of different types can be advantageous. A particularly interesting solution is obtained for the case where the array contains an equal number of sensors, shown in figure \ref{fig:arraygeometries}d), where all sensors cluster near the $x$-axis. In a GW detector, this area coincides with one of the interferometer arms, which are several kilometers long. Such an array could be installed inside of the arm infrastructure, reducing the need for borehole deployments.

\begin{figure}[!htbp]
    \centering
    \includegraphics[width=0.6\linewidth]{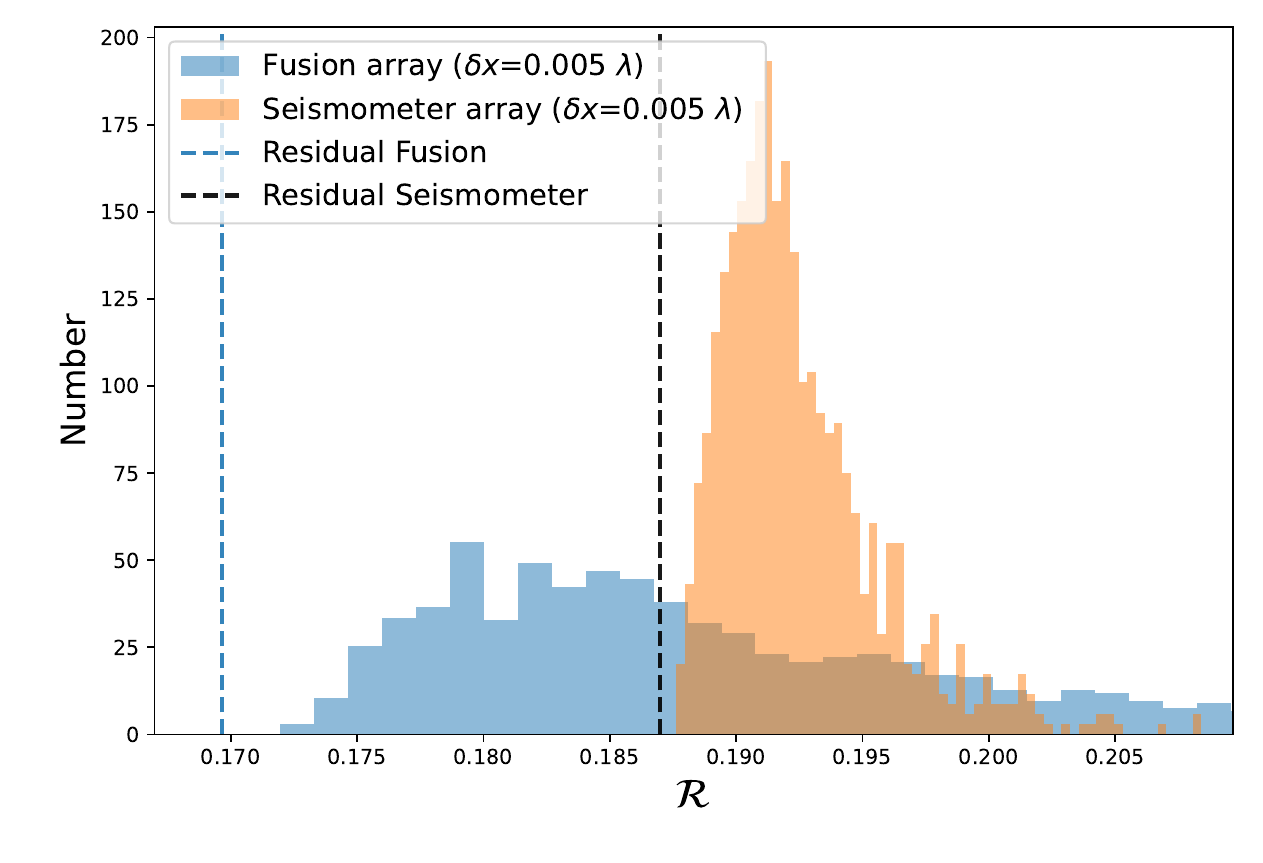}
    \caption{Distribution of 1000 WF filter residuals $\mathcal{R}$ for sensor arrays using a Monte Carlo simulation. Each sensor coordinate was perturbed by an independent offset drawn from a normal distribution with standard deviation $\sigma = 0.005 \lambda$, and starting initially from an optimized configuration. The figure compares a fusion array with equal numbers of seismometers and strainmeters ($N_{\mathrm{seis}} = N_{\mathrm{strain}} = 3$, labeled “Fusion”) and a seismometer-only array ($N_{\mathrm{seis}} = 6$, $N_{\mathrm{strain}} = 0$, labeled “Seismometer”). The dashed lines show the optimized residual.}
    \label{fig:robustness}
\end{figure}

We further tested the robustness of the optimized solutions against positional perturbations, following the approach in  \cite{harms_optimization_2019}. To assess the robustness of the optimized configuration, we performed a Monte-Carlo perturbation test by adding random shifts drawn from a normal distribution with a standard deviation of $0.005,\lambda$ (e.g., $2.5$ m for a wavelength of $500$ m) to all sensor coordinates. This procedure quantifies the WF filter performance loss due to sensor misplacement. We performed 1000 realizations and compared the resulting histograms of residuals for two cases: (i) a fusion array with equal numbers of strainmeters and seismometers ($N_{\mathrm{seis}} = N_{\mathrm{strain}} = 3$), and (ii) a seismometer-only array (Fig.~\ref{fig:robustness}). The results reveal a difference in variability: the fusion array exhibits a residual spread of $\sigma = 1.22 \times 10^{-2}$, whereas the seismometer array shows a narrower distribution with $\sigma = 3.16 \times 10^{-3}$. This indicates that the seismometer-only configuration is less sensitive to sensor misplacement. However, DAS could mitigate this limitation, since its dense deployment capabilities of about one sensor per meter may help compensate for position mismatches. 

\begin{figure}[!htbp]
    \centering
    \includegraphics[width=0.7\linewidth]{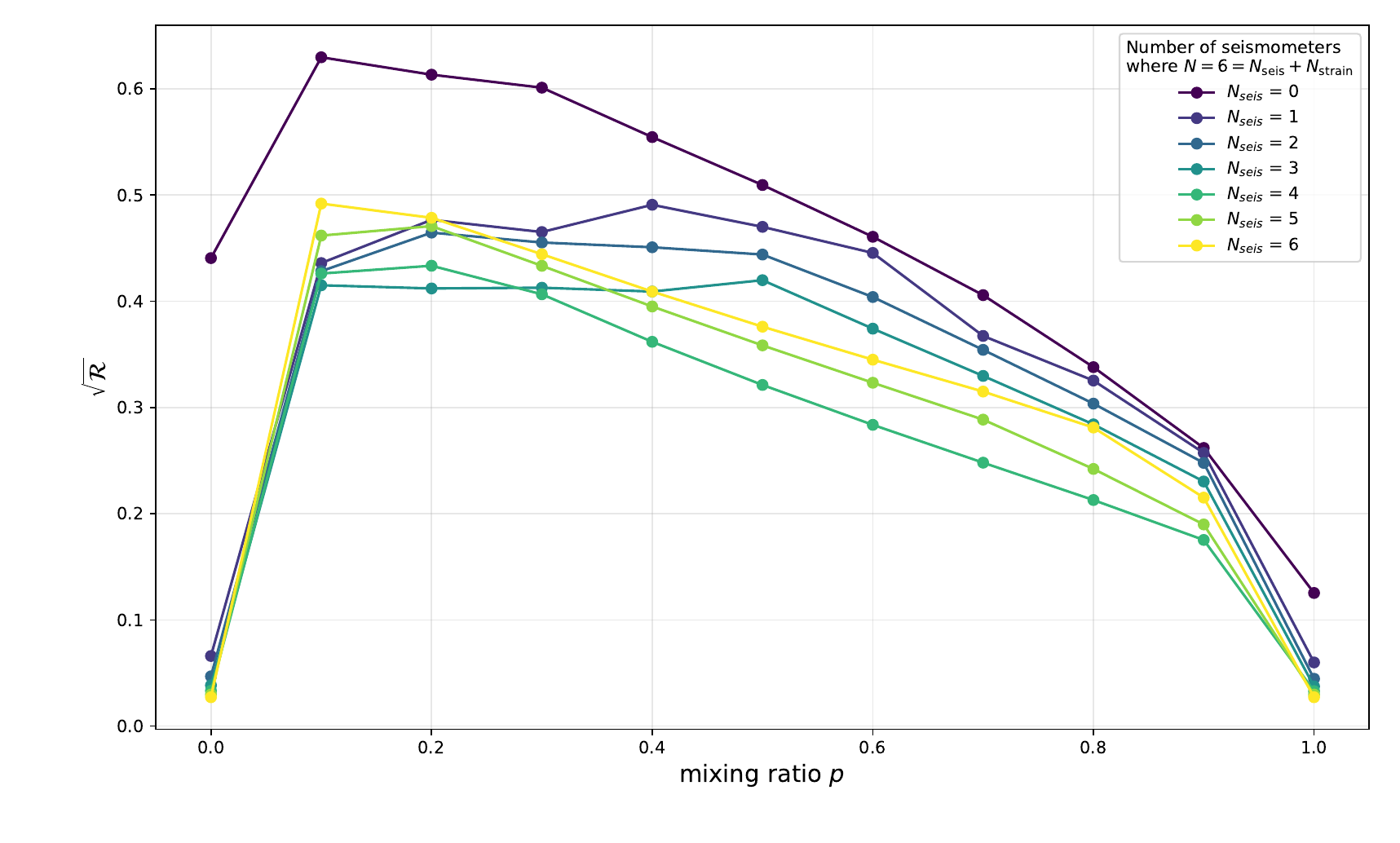}
    \caption{This figure shows the dependence of the square root of the optimal WF filter residual $\sqrt{\mathcal{R}}$, on the mixing ratio $p$ between P- and S-waves. The number of seismometers ($N_{\mathrm{seis}}$) in a fusion array composed of 6 sensors in total, where the remaining sensors are strainmeters ($N_{\mathrm{strain}}$), is color coded. All sensors are assumed to have an SNR of 15, and their one axis measurement directions are aligned with that of the TM. }
    \label{fig:mixingp}
\end{figure}

On top of testing the robustness against positional shifts, we also investigated how the solutions change when varying the mixing ratio of the seismic field’s P- and S-wave content. This is shown in figure \ref{fig:mixingp}. For $p=0$ and $p=1$, all arrays containing at least one seismometer achieve almost perfect cancellation. This is, however, not the case for strainmeter-only arrays, which cannot simply be placed next to the TM and still provide good correlations. In particular, when the seismic noise field is composed exclusively of S-waves, the achievable cancellation with strainmeters is very low, since $\sqrt{\mathcal{R}} = 0.63$. This originates from the fact that strainmeter correlations are generally stronger for P-waves than for S-waves, as can be read off the prefactors of the correlation coefficients in table \ref{tab:correlation_coefficients_explicit}. Otherwise, for all sensor arrays, the slope shows a similar behavior in dependence on the p value: Minimal residuals at p = 0 and p=1, but worsen performance for mixed seismic wave fields. 
 
\subsection{Fusion Sensor Arrays: Capabilities of NN Cancellation}\label{sec: FA evaluation}

\begin{figure}[!htbp]
    \centering
    \includegraphics[width=0.5\linewidth]{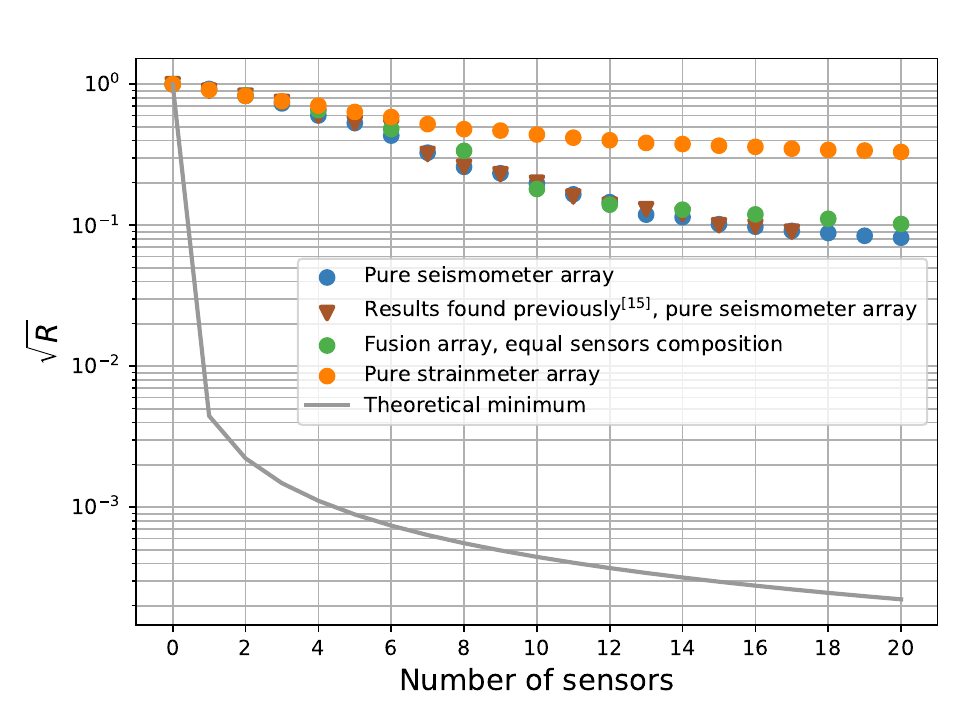}
    \caption{This figure shows the dependence of the square root of the WF filter residual $\sqrt{\mathcal{R}}$ on the number of sensors $N$, for optimized sensor arrays. Results are given for arrays composed exclusively of strainmeters, exclusively of seismometers, and for fusion arrays with equal numbers of both sensor types. The theoretically best possible performance of the WF for a seismometer array is given by $1/(\mathrm{SNR}^2 \cdot N)$, which was derived in  \cite{harms_optimization_2019} and is shown as a gray line. We further compare the results for the seismometer-only array to those obtained in \cite{harms_optimization_2019}. The parameters of the seismic field and sensors are the same as those described in section \ref{sec: Validation}.}
    \label{fig:residualvsN}
\end{figure}

Having finished the validation and first evaluation of fusion sensor arrays, we continue by analyzing on how to achieve the NN cancellation factors of the order of 10, which are necessary to reach the design sensitivity of ET \cite{Amann:2022pyq}. Therefore, we analyze the performance of seismometer, strainmeter and fusion sensor arrays as a function of the total number of sensors $N$ in the array and use up to 20 sensors. The results are shown in figure \ref{fig:residualvsN}. We focus on fusion arrays composed by an equal number of seismometers and strainmeters, because this configuration was found to provide the optimal composition for a total of 6 sensors, see figure \ref{fig:N20compositionandET}. Apart from the number of sensors, we use the same setup for the seismic field and sensors SNR and measurement direction, as described in section \ref{sec: Validation}.

The performance of the seismometer arrays shown in figure \ref{fig:residualvsN} agrees with previous results for single-axis seismometers \cite{harms_optimization_2019}. In contrast, the strainmeter arrays perform worse than the other configurations, consistent with the results discussed in section \ref{sec: Validation}. Optimized fusion arrays perform similarly to seismometer arrays across the entire range of $N$, hence that they can compete with seismometer-only arrays. 
 \\
 % Fitting the optimized ET-constrained values with an exponential function, $f(x) = a \, e^{-b \cdot x + c}$, we find that a square root WF filter residual of $0.1$ would be reached with 128 sensors (64 seismometers and 64 strainmeters). Based on this estimate, we optimized a array with 130 sensors, consisting of 65 seismometers and 65 strainmeters, and obtained, without much fine-tuning, a residual of $\sqrt{\mathcal{R}} = 0.0407$. Whether such a solution can be realized in practice depends strongly on the specific ET infrastructure. Moreover, our analysis was restricted to a particular setup and a relatively short wavelength of $500 \,\mathrm{m}$. Nevertheless, we conclude that in certain scenarios, where a sensor array is installed in alignment with the ET infrastructure, a fusion array may provide advantages over arrays composed exclusively of strainmeters or seismometers.

\subsection{Sensor arrays in the vicinity of the ET infrastructure}\label{sec: FA in ET}
Since positioning sensors optimally for NN mitigation typically involves drilling costly boreholes, it is natural to consider alternative configurations located near the ET infrastructure. The optimizations performed so far aimed at finding the global minimum of the WF residual function, without constraints on the sensor positions. We now seek to determine the minimum defined by equation \eqref{eq: minimization problem} under the additional constraint that the solution space is restricted to positions near the infrastructure.

\begin{figure}[!htbp]
    \centering
    \includegraphics[width=0.5\linewidth]{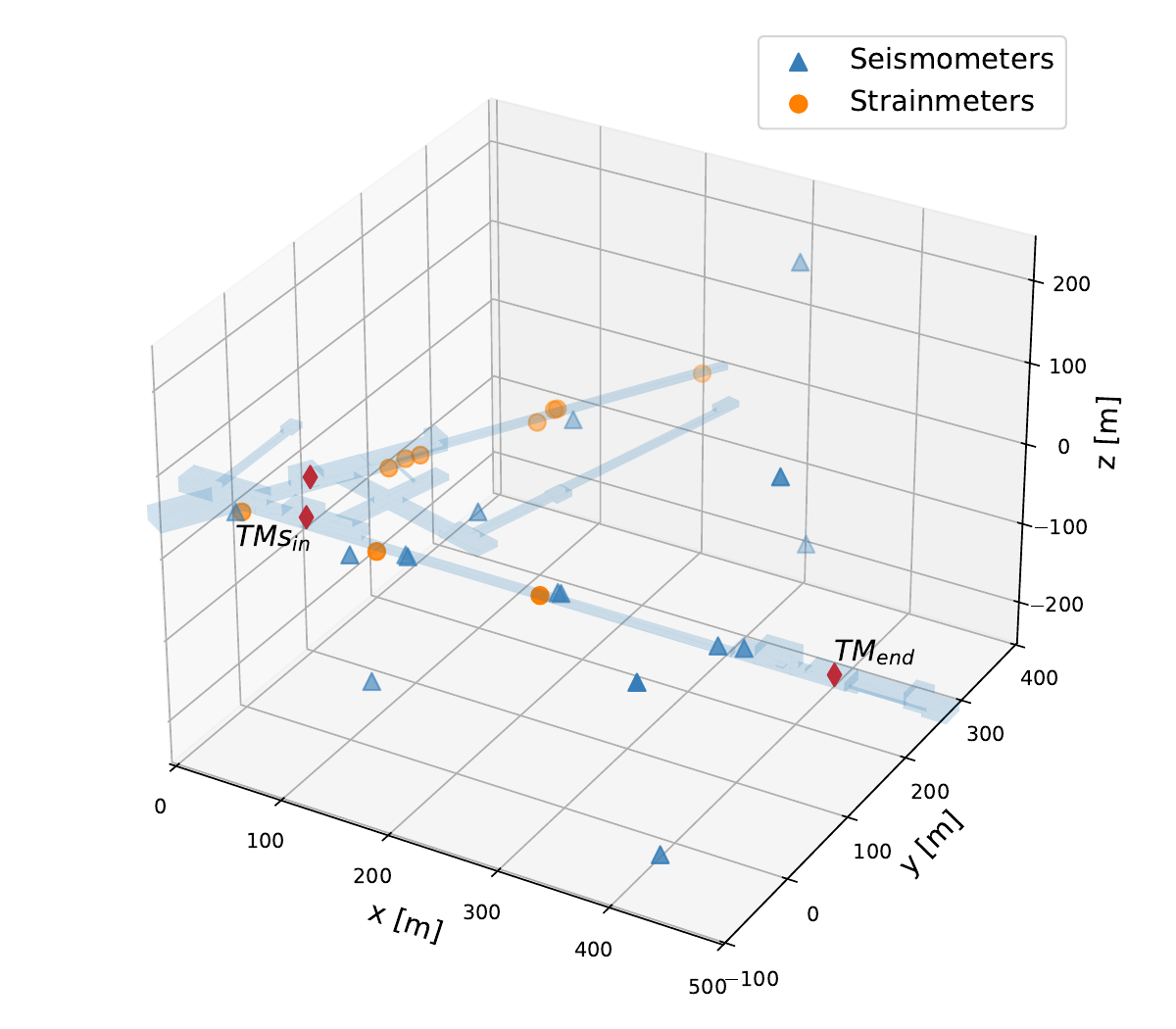}
    \caption{This figure shows an example array configuration, where the seismometers are placed around the ET infrastructure and the strainmeters are placed inside of the ET infrastructure. The '$TMs_{in}$' indicates the input and '$TM_{end}$' the end mirrors in the Michelson interferometer. The design (blue shaded area) is from the optical layout \cite{Optical_layout_ET}. We include multiple TMs only for illustrative purposes, the optimization is performed for one single TM.}
    \label{fig:ET_layout_constrainedarray}
\end{figure}

In the DE and CMA-ES optimization algorithms, the search domain is a required parameter, which makes them well suited for this analysis. For the present study, we used a box of size $\mathcal{B} = [-500,500] \times [-20,20] \times [-20,20] \, \mathrm{m}^3$ and considered a seismic wavelength of $500 \, \mathrm{m}$. The Box $\mathcal{B}$ approximates the volume covered by the tunnel of an ET arm. While the domain $\mathcal{B}$ may exceed the actual extent of the ET infrastructure by a few meters, depending on the final design of ET, it provides a useful indication of the achievable performance and a lower bound for the residuals. The updated optimization problem reads
\begin{equation} \label{eq: minimization problem}
    \mathcal{R}(\{{\mathbf{x}_0\}_i) \leq \mathcal{R}(\{{\mathbf{x}\}}_i)}, \space \forall \{{\mathbf{x}\}}_{i\in J} \in (\mathbb{R}^3)^N,\{{\mathbf{x}\}}_{i\in K} \in \mathcal{B},
\end{equation}
where $J$ and $K$ are index sets, which distribute the sensor positions either to the full space or the ET arm.

\begin{figure}[!htbp]
    \centering
    \includegraphics[width=0.5\linewidth]{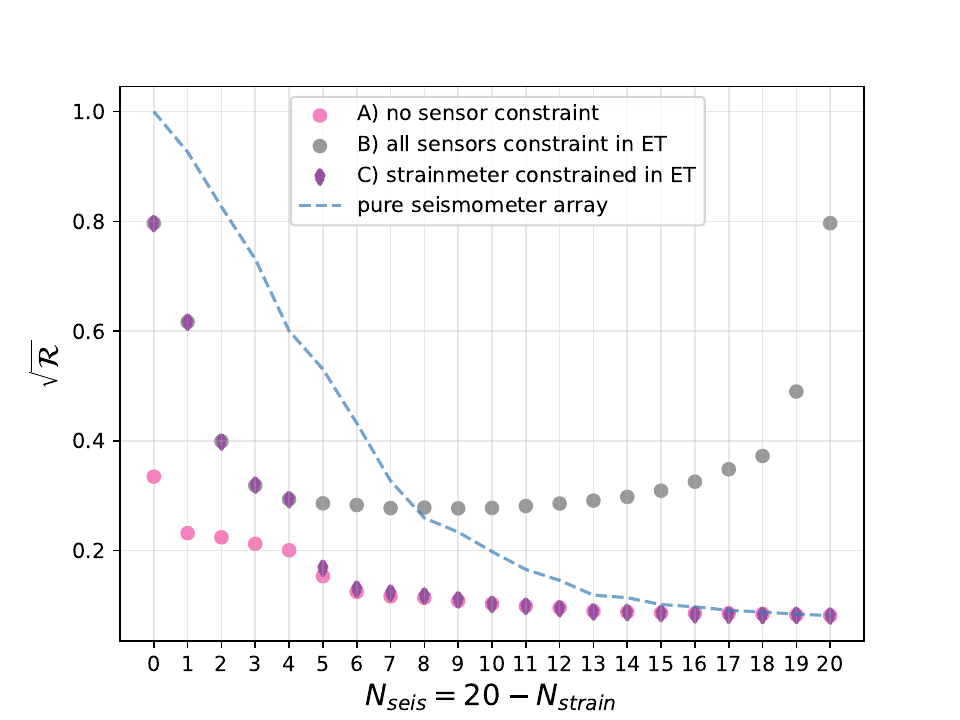}
    \caption{This figure presents the analysis of the optimal composition for a fusion sensor array composed of $N_{\text{seis}}$ seismometers and $N_{\text{strain}}$ strainmeters, with a total of $N = 20$ sensors. The y-ordinate shows the square root of the WF residual is shown, and the x-ordinate, the number of seismometers. The pink points are not constrained, the grey points show the results for arrays which are totally constrained to the area $\mathcal{B} = [-500,500] \times [-20,20] \times [-20,20] \, \mathrm{m}^3$ and the purple points are solutions where only the strainmeters are constrained to $\mathcal{B}$. The dashed blue line shows the performance of arrays with no strainmeters, i.e., have a total of $N = N_{\text{seis}}$ sensors. The seismic field has a mixing ratio of $p = 1/3$, and the SNR of all sensors is given by SNR$ = 15$.}
    \label{fig:N20compositionandET}
\end{figure}

We investigate the NN mitigation capabilities of sensor arrays partly installed in the ET infrastructure by analyzing the WF residual for all of the different compositions of strain- and seismometers for an array composed of 20 sensors in total. We analyze the cases:
\begin{itemize}
    \item A) an array that is not constrained to $\mathcal{B}$ at all.
    \item B) an array which is totally constrained to the tunnel of the ET arm $\mathcal{B}$.
    \item C) an array where only the strainmeters are constrained to $\mathcal{B}$.
\end{itemize} 
The results are shown in figure \ref{fig:N20compositionandET}. 

The results for case A) shown in figure \ref{fig:N20compositionandET} as pink points can be compared to the composition analysis shown in figure \ref{fig:compositionfusionarrayN6}. The slope of the two is different, the minimum for $N = 20$ is attained at $N_{\text{seis}} = 20$, where all sensors in the fusion array are seismometers. This is in contrast to the case where we had $N = 6$ sensors in total, where the minimum was attained using an equal composition of seismometer and strainmeter. This stems from the fact that, for a sensor array of 20 sensors, the main limitation does not originate from the $p$ mixing ratio, but rather from the SNR of the sensors. Fusion sensor arrays are not as effective for higher number sensor networks as strainmeter sensors are not as sensitive to NN naturally. However, as shown in figure \ref{fig:N20compositionandET}, the difference between $\sqrt{\mathcal{R}}(N_{\text{seis}} = 6)$ and $\sqrt{\mathcal{R}}(N_{\text{seis}} = 20)$ is only $\approx 2 \%$.

The results for case B), for an array completely installed inside of the ET arm, show that mixing seismometers and strainmeters is indeed beneficial. The minimum of $\sqrt{R} = 0.277$ is found for $N_{\text{seis}} = 7$, hence for a nearly equal number of sensors for each type. Further, the mono type sensor optimized arrays perform poorly compared to the fusion arrays, where for both sensor types $\sqrt{R} = 0.8$ is found. Constraining all sensors to the ET infrastructure, the performance is worse compared to unconstrained case A) for all array configurations, and the best results found are $\sqrt{R} = 0.277$ for B), compared to $\sqrt{R} = 0.08155746$ for A).

The results for case C), where only the strainmeters are constrained to ET, show a very interesting pattern, as they start to be equal to the results from case B) where all sensors were constrained for $N_{\text{seis}} \leq 4$, but for $N_{\text{seis}} > 4$ the results are equal to the residual found in case A), where no sensor is constrained. It seems that the position of the strainmeter is not much relevant for the performance of fusion sensor arrays, once a certain number of seismometers is given. And since the performance for the fusion sensor arrays $\sqrt{\mathcal{R}}(N_{\text{seis}} = 20)$ does not differ much from the performance $\sqrt{\mathcal{R}}(N_{\text{seis}} = 6)$, the number of seismometers in boreholes could be set to 6 and by using 14 additional strainmeters within ET infrastructure a performance of $\sqrt{R} = 0.1248$, equivalent to the performance for 20 seismometers in boreholes, given by $\sqrt{R} = 0.08155746$, can be achieved. This is an improvement of a factor of approximately 4 in square root residual, hence NN left in the signal, compared to a pure seismometer array (blue dashed line in figure \ref{fig:N20compositionandET}) with 6 sensors.

\section{Conclusions}

This paper builds on previous studies that analyzed the capabilities of sensor arrays to mitigate Newtonian noise (NN) for the Einstein Telescope (ET) \cite{harms_optimization_2019, badaracco_joint_2024}. We introduced strain correlation coefficients for S-waves and investigated the NN-cancellation performance of sensor-fusion arrays. Furthermore, we employed a new optimization framework combining Differential Evolution and the Covariant Matrix Adaptation Evolution Strategy.
We first evaluated arrays composed of both seismometers and strainmeters. Since we introduced a new Wiener filter (WF) formalism for mixed sensor types, we benchmarked our results against previous findings and analyzed them in detail. We subsequently investigated arrays that, rather than being installed in separate boreholes, are located in the vicinity of the ET infrastructure. Both approaches have the potential to significantly reduce the overall installation costs of a NN cancellation system for ET.
We then examined fusion arrays consisting of a total of six sensors. Our results show that combining different sensor types can be advantageous for NN cancellation, as their complementary information enhances overall performance. For arrays with a low number of total sensors, where the mixing ratio $p$ limits the WF performance, this leads to residuals that are comparable to those of seismometer-only arrays, or improved by a few percent. The optimized configurations of fusion arrays display a variety of geometries.

% We also assess the robustness of these solutions with respect to positional misplacements and variations in the seismic field mixing ratio, thereby exploring the parameter space of our model. Overall, fusion sensor arrays exhibit behavior similar to that of seismometer arrays with respect to these parameters. Extending our analysis to larger arrays containing up to 20 sensors, we find that fusion sensor arrays achieve performance levels comparable to traditional seismometer arrays even at higher sensor counts.

We evaluated fusion sensor arrays installed fully or partially within the ET infrastructure. Arrays that are entirely constrained to the ET geometry show WF residuals that are, on average, worse by a factor of approximately three compared to unconstrained configurations. Our results indicate that the optimal composition of fusion arrays consists of an approximately equal number of seismometers and strainmeters. We further examined a configuration in which only the strainmeters are constrained to the ET infrastructure. This setup achieves performance comparable to that of the unconstrained array when at least five seismometers are included, suggesting that 14 external boreholes could be replaced by an optimized strainmeter array installed within ET. In summary, we demonstrated that fusion sensor arrays offer several advantages for NN cancellation.

The findings of the results are limited by several simplified assumptions. The here employed analytical setup assumes a stationary, isotropic, uncorrelated, and monochromatic seismic noise floor. Further, the calculations are done in full space, which is not realistic, as the depth of ET is not larger than the average seismic wavelength. In fact, using realistic seismic wave speeds of $v = 5000$ m/s and the frequencies which are of interest for NN, e.g., 10 Hz, the wavelength of $500$ m is twice as large as the proposed depth for ET of $250$ m, thus our assumption of a full space is invalid. The medium surrounding ET will also not be homogeneous, as assumed in the analysis. Further, we only considered body waves for the seismic wave fields, assuming that Rayleigh waves are attenuated already at these depths, which is an assumption which needs to be validated by the borehole campaigns of the respective candidate sites for ET. Most of these assumptions can only be relaxed using simulations for the seismic field, which would require a very different level of analysis. Also the optimization algorithms have to be adapted. While the isotropic case can be considered somewhat of a worst case scenario for the NN cancellation performance, this is generally more complicated, e.g., going from full to half space, which is why better models for seismic NN are needed.

In this work, we analyzed the advantages of disentangling the P- and S-wave content by using two sensor types, which react differently to the different wave types. Another advantage of strainmeter sensors, including DAS, is that they can be installed densely yet still cover large regions. Therefore, there are more advantages of these sensors to be explored, such as exploring the WF capabilities of mitigating NN for a large number of sensors, or capturing full wavefront and then modeling the wave field in more advanced ways, including a time domain. Further, the NN cancellation, averaged over portions of the parameter space (e.g., frequency), should be improved, as many different configurations can be included in one installation, compared to sensor arrays, which are limited in sensor number.

\newpage

\section*{Acknowledgements}
The authors thank Jan Harms for many useful discussions regarding the Wiener filter and strainmeters. We also thank Patrick Schillings for discussions on the optimization algorithm and for providing an implementation of ADAM (publicly available; see the ``Availability of Code'' section in \cite{schillings_fighting_2025}), although it was not used in the final analysis. We further thank Jonathan Bratanata for providing the IFC file used to illustrate the ET optical design in Figure~\ref{fig:ET_layout_constrainedarray}.

\section*{Funding}
We acknowledge Helmut-Schmidt-University and DESY (members of the Helmholtz Association), as well as DASHH---Data Science in Hamburg, Helmholtz Graduate School for the Structure of Matter---for financial support.

Computational resources (HPC cluster HSUper) were provided via the project hpc.bw, funded by dtec.bw---Digitalization and Technology Research Center of the Bundeswehr. dtec.bw is funded by the European Union---NextGenerationEU.

\section*{Data and code availability}
The code required to reproduce the results is available at \url{https://github.com/PaulOphardt31/Positions_Optimization_fusion_sensor_arrays/tree/main}.

\bibliographystyle{unsrt}
\bibliography{FNNC}

@article{Amann:2022pyq,
    author = "Amann, Florian and Badaracco, Francesca and DeSalvo, Riccardo and Naticchioni, Luca and Paoli, Andrea and Paoli, Luca and Ruggi, Paolo and Selleri, Stefano",
    title = "{Tunnel Configurations and Seismic Isolation Optimization in Underground Gravitational Wave Detectors}",
    eprint = "2204.04131",
    archivePrefix = "arXiv",
    primaryClass = "astro-ph.IM",
    doi = "10.3390/app12178827",
    journal = "Appl. Sciences",
    volume = "12",
    number = "17",
    pages = "8827",
    year = "2022"
}

@misc{Optical_layout_ET,
  author       = {{Einstein Telescope Collaboration}},
  title        = {Einstein Telescope optical layout (Trimble Connect model)},
  howpublished = {\url{https://web.connect.trimble.com/projects/bgr1skWvm0U/viewer/3d/?modelId=F8Gg70Iz6ik,mQSxHbI-jIc&=&origin=app21.connect.trimble.com&stoken=FzrrAFZw088AKW5uuVLQ_G7fZhN78-3_z2rt5_e2LkaLb8lmGndsjFTWhM3PpeEp}},
  note         = {Accessed: 2025-11-07},
  year         = {2025}
}

@article{lu_distributed_2019,
	title = {Distributed optical fiber sensing: {Review} and perspective},
	volume = {6},
	shorttitle = {Distributed optical fiber sensing},
	url = {https://aip.scitation.org/doi/abs/10.1063/1.5113955},
	doi = {10.1063/1.5113955},
	number = {4},
	urldate = {2020-06-15},
	journal = {Applied Physics Reviews},
	author = {Lu, Ping and Lalam, Nageswara and Badar, Mudabbir and Liu, Bo and Chorpening, Benjamin T. and Buric, Michael P. and Ohodnicki, Paul R.},
	month = oct,
	year = {2019},
	pages = {041302},
}

@article{wang_recent_2020,
	title = {Recent {Progress} in {Distributed} {Fiber} {Acoustic} {Sensing} with phi-{OTDR}},
	volume = {20},
	copyright = {http://creativecommons.org/licenses/by/3.0/},
	url = {https://www.mdpi.com/1424-8220/20/22/6594},
	doi = {10.3390/s20226594},
	language = {en},
	number = {22},
	urldate = {2021-01-22},
	journal = {Sensors},
	author = {Wang, Zhaoyong and Lu, Bin and Ye, Qing and Cai, Haiwen},
	month = jan,
	year = {2020},
	note = {Number: 22
Publisher: Multidisciplinary Digital Publishing Institute},
	pages = {6594},
}

@article{zinsou_recent_2019,
	title = {Recent {Progress} in the {Performance} {Enhancement} of {Phase}-{Sensitive} {OTDR} {Vibration} {Sensing} {Systems}},
	volume = {19},
	copyright = {http://creativecommons.org/licenses/by/3.0/},
	url = {https://www.mdpi.com/1424-8220/19/7/1709},
	doi = {10.3390/s19071709},
	language = {en},
	number = {7},
	urldate = {2021-02-02},
	journal = {Sensors},
	author = {Zinsou, Romain and Liu, Xin and Wang, Yu and Zhang, Jianguo and Wang, Yuncai and Jin, Baoquan},
	month = jan,
	year = {2019},
	note = {Number: 7
Publisher: Multidisciplinary Digital Publishing Institute},
	pages = {1709},
}

@article{miah_review_2017,
	title = {A {Review} of {Hybrid} {Fiber}-{Optic} {Distributed} {Simultaneous} {Vibration} and {Temperature} {Sensing} {Technology} and {Its} {Geophysical} {Applications}},
	volume = {17},
	copyright = {http://creativecommons.org/licenses/by/3.0/},
	url = {https://www.mdpi.com/1424-8220/17/11/2511},
	doi = {10.3390/s17112511},
	language = {en},
	number = {11},
	urldate = {2021-01-13},
	journal = {Sensors},
	author = {Miah, Khalid and Potter, David K.},
	month = nov,
	year = {2017},
	note = {Number: 11
Publisher: Multidisciplinary Digital Publishing Institute},
	pages = {2511},
}

@article{yuan_waveform_2024,
	title = {Waveform reconstruction of core-collapse supernova gravitational waves with ensemble empirical mode decomposition},
	volume = {529},
	copyright = {https://creativecommons.org/licenses/by/4.0/},
	issn = {0035-8711, 1365-2966},
	url = {https://academic.oup.com/mnras/article/529/4/3235/7634213},
	doi = {10.1093/mnras/stae604},
	language = {en},
	number = {4},
	urldate = {2025-11-04},
	journal = {Monthly Notices of the Royal Astronomical Society},
	author = {Yuan, Yong and Fan, Xi-Long and Lü, Hou-Jun and Sun, Yang-Yi and Lin, Kai},
	month = mar,
	year = {2024},
	pages = {3235--3243},
}

@article{sieniawska_gravitational_2021,
	title = {Gravitational waves from spinning neutron stars as not-quite-standard sirens},
	volume = {509},
	copyright = {https://academic.oup.com/journals/pages/open\_access/funder\_policies/chorus/standard\_publication\_model},
	issn = {0035-8711, 1365-2966},
	url = {https://academic.oup.com/mnras/article/509/4/5179/6430870},
	doi = {10.1093/mnras/stab3315},
	language = {en},
	number = {4},
	urldate = {2025-11-04},
	journal = {Monthly Notices of the Royal Astronomical Society},
	author = {Sieniawska, Magdalena and Jones, David Ian},
	month = dec,
	year = {2021},
	pages = {5179--5187},
}

@article{maggiore_science_2019,
	title = {Science {Case} for the {Einstein} {Telescope}},
    journal = {J. Cosmol. Astropart. Phys.},
	copyright = {arXiv.org perpetual, non-exclusive license},
	url = {https://arxiv.org/abs/1912.02622},
	doi = {10.48550/ARXIV.1912.02622},
	urldate = {2025-11-04},
	author = {Maggiore, Michele and Broeck, Chris van den and Bartolo, Nicola and Belgacem, Enis and Bertacca, Daniele and Bizouard, Marie Anne and Branchesi, Marica and Clesse, Sebastien and Foffa, Stefano and García-Bellido, Juan and Grimm, Stefan and Harms, Jan and Hinderer, Tanja and Matarrese, Sabino and Palomba, Cristiano and Peloso, Marco and Ricciardone, Angelo and Sakellariadou, Mairi},
	year = {2019},
	note = {Publisher: arXiv
Version Number: 4},
}

@article{koehn_impact_2024,
	title = {Impact of dark matter on tidal signatures in neutron star mergers with the {Einstein} {Telescope}},
	volume = {110},
	issn = {2470-0010, 2470-0029},
	url = {https://link.aps.org/doi/10.1103/PhysRevD.110.103033},
	doi = {10.1103/PhysRevD.110.103033},
	language = {en},
	number = {10},
	urldate = {2025-11-04},
	journal = {Physical Review D},
	author = {Koehn, Hauke and Giangrandi, Edoardo and Kunert, Nina and Somasundaram, Rahul and Sagun, Violetta and Dietrich, Tim},
	month = nov,
	year = {2024},
	pages = {103033},
}

@article{pagliaro_searching_2025,
	title = {Searching for continuous gravitational waves from slowly spinning neutron stars with {DECIGO}, {Big} {Bang} {Observer}, {Einstein} {Telescope}, and {Cosmic} {Explorer}},
	volume = {540},
	copyright = {https://creativecommons.org/licenses/by/4.0/},
	issn = {0035-8711, 1365-2966},
	url = {https://academic.oup.com/mnras/article/540/1/1006/8128264},
	doi = {10.1093/mnras/staf774},
	language = {en},
	number = {1},
	urldate = {2025-11-04},
	journal = {Monthly Notices of the Royal Astronomical Society},
	author = {Pagliaro, Gianluca and Papa, Maria Alessandra and Ming, Jing and Muratore, Martina},
	month = may,
	year = {2025},
	pages = {1006--1016},
}

@article{vartanyan_gravitational_2020,
	title = {Gravitational {Waves} from {Neutrino} {Emission} {Asymmetries} in {Core}-collapse {Supernovae}},
	volume = {901},
	issn = {0004-637X, 1538-4357},
	url = {https://iopscience.iop.org/article/10.3847/1538-4357/abafac},
	doi = {10.3847/1538-4357/abafac},
	number = {2},
	urldate = {2025-11-04},
	journal = {The Astrophysical Journal},
	author = {Vartanyan, David and Burrows, Adam},
	month = oct,
	year = {2020},
	pages = {108},
}

@misc{giangrandi_numerical_2025,
	title = {Numerical {Relativity} {Simulations} of {Dark} {Matter} {Admixed} {Binary} {Neutron} {Stars}},
	copyright = {Creative Commons Attribution 4.0 International},
	url = {https://arxiv.org/abs/2504.20825},
	doi = {10.48550/ARXIV.2504.20825},
	urldate = {2025-11-04},
	publisher = {arXiv},
	author = {Giangrandi, Edoardo and Rüter, Hannes R. and Kunert, Nina and Emma, Mattia and Abac, Adrian and Adhikari, Ananya and Dietrich, Tim and Sagun, Violetta and Tichy, Wolfgang and Providência, Constança},
	year = {2025},
	note = {Version Number: 2},
}

@article{punturo_einstein_2010,
	title = {The {Einstein} {Telescope}: a third-generation gravitational wave observatory},
	volume = {27},
	issn = {0264-9381, 1361-6382},
	shorttitle = {The {Einstein} {Telescope}},
	url = {https://iopscience.iop.org/article/10.1088/0264-9381/27/19/194002},
	doi = {10.1088/0264-9381/27/19/194002},
	number = {19},
	urldate = {2025-11-04},
	journal = {Classical and Quantum Gravity},
	author = {Punturo, M and Abernathy, M and Acernese, F and Allen, B and Andersson, N and Arun, K and Barone, F and Barr, B and Barsuglia, M and Beker, M and Beveridge, N and Birindelli, S and Bose, S and Bosi, L and Braccini, S and Bradaschia, C and Bulik, T and Calloni, E and Cella, G and Mottin, E Chassande and Chelkowski, S and Chincarini, A and Clark, J and Coccia, E and Colacino, C and Colas, J and Cumming, A and Cunningham, L and Cuoco, E and Danilishin, S and Danzmann, K and De Luca, G and De Salvo, R and Dent, T and De Rosa, R and Di Fiore, L and Di Virgilio, A and Doets, M and Fafone, V and Falferi, P and Flaminio, R and Franc, J and Frasconi, F and Freise, A and Fulda, P and Gair, J and Gemme, G and Gennai, A and Giazotto, A and Glampedakis, K and Granata, M and Grote, H and Guidi, G and Hammond, G and Hannam, M and Harms, J and Heinert, D and Hendry, M and Heng, I and Hennes, E and Hild, S and Hough, J and Husa, S and Huttner, S and Jones, G and Khalili, F and Kokeyama, K and Kokkotas, K and Krishnan, B and Lorenzini, M and Lück, H and Majorana, E and Mandel, I and Mandic, V and Martin, I and Michel, C and Minenkov, Y and Morgado, N and Mosca, S and Mours, B and Müller–Ebhardt, H and Murray, P and Nawrodt, R and Nelson, J and Oshaughnessy, R and Ott, C D and Palomba, C and Paoli, A and Parguez, G and Pasqualetti, A and Passaquieti, R and Passuello, D and Pinard, L and Poggiani, R and Popolizio, P and Prato, M and Puppo, P and Rabeling, D and Rapagnani, P and Read, J and Regimbau, T and Rehbein, H and Reid, S and Rezzolla, L and Ricci, F and Richard, F and Rocchi, A and Rowan, S and Rüdiger, A and Sassolas, B and Sathyaprakash, B and Schnabel, R and Schwarz, C and Seidel, P and Sintes, A and Somiya, K and Speirits, F and Strain, K and Strigin, S and Sutton, P and Tarabrin, S and Thüring, A and Van Den Brand, J and Van Leewen, C and Van Veggel, M and Van Den Broeck, C and Vecchio, A and Veitch, J and Vetrano, F and Vicere, A and Vyatchanin, S and Willke, B and Woan, G and Wolfango, P and Yamamoto, K},
	month = oct,
	year = {2010},
	pages = {194002},
}

@article{harms_terrestrial_2019,
	title = {Terrestrial gravity fluctuations},
	volume = {22},
	issn = {2367-3613, 1433-8351},
	url = {http://link.springer.com/10.1007/s41114-019-0022-2},
	doi = {10.1007/s41114-019-0022-2},
	language = {en},
	number = {1},
	urldate = {2025-11-04},
	journal = {Living Reviews in Relativity},
	author = {Harms, Jan},
	month = dec,
	year = {2019},
	pages = {6},
}

@misc{harms_lower_2022,
	title = {A lower limit for {Newtonian}-noise models of the {Einstein} {Telescope}},
	copyright = {arXiv.org perpetual, non-exclusive license},
	url = {https://arxiv.org/abs/2202.12841},
	doi = {10.48550/ARXIV.2202.12841},
	urldate = {2025-11-04},
	publisher = {arXiv},
	author = {Harms, Jan and Naticchioni, Luca and Calloni, Enrico and De Rosa, Rosario and Ricci, Fulvio and D'Urso, Domenico},
	year = {2022},
	note = {Version Number: 1},
}

@article{singha_newtonian-noise_2020,
	title = {Newtonian-noise reassessment for the {Virgo} gravitational-wave observatory including local recess structures},
	volume = {37},
	issn = {0264-9381, 1361-6382},
	url = {https://iopscience.iop.org/article/10.1088/1361-6382/ab81cb},
	doi = {10.1088/1361-6382/ab81cb},
	number = {10},
	urldate = {2025-11-04},
	journal = {Classical and Quantum Gravity},
	author = {Singha, Ayatri and Hild, Stefan and Harms, Jan},
	month = may,
	year = {2020},
	pages = {105007},
}

@article{koley_design_2024,
	title = {Design and implementation of a seismic {Newtonian} noise cancellation system for the {Virgo} gravitational-wave detector},
	volume = {139},
	issn = {2190-5444},
	url = {https://link.springer.com/10.1140/epjp/s13360-023-04834-0},
	doi = {10.1140/epjp/s13360-023-04834-0},
	language = {en},
	number = {1},
	urldate = {2025-11-04},
	journal = {The European Physical Journal Plus},
	author = {Koley, Soumen and Harms, Jan and Allocca, Annalisa and Badaracco, Francesca and Bertolini, Alessandro and Bulik, Tomasz and Calloni, Enrico and Cieslar, Marek and De Rosa, Rosario and Errico, Luciano and Esposito, Marina and Fiori, Irene and Hild, Stefan and Idzkowski, Bartosz and Masserot, Alain and Mours, Benoît and Paoletti, Federico and Paoli, Andrea and Pietrzak, Mateusz and Rei, Luca and Rolland, Loïc and Singha, Ayatri and Suchenek, Mariusz and Suchinski, Maciej and Tringali, Maria Concetta and Ruggi, Paolo},
	month = jan,
	year = {2024},
	pages = {48},
}

@article{harms_passive_2014,
	title = {Passive {Newtonian} noise suppression for gravitational-wave observatories based on shaping of the local topography},
	volume = {31},
	issn = {0264-9381, 1361-6382},
	url = {http://arxiv.org/abs/1406.2253},
	doi = {10.1088/0264-9381/31/18/185011},
	number = {18},
	urldate = {2025-11-04},
	journal = {Classical and Quantum Gravity},
	author = {Harms, Jan and Hild, Stefan},
	month = sep,
	year = {2014},
	note = {arXiv:1406.2253 [gr-qc]},
	pages = {185011},
}

@article{driggers_subtraction_2012,
	title = {Subtraction of {Newtonian} noise using optimized sensor arrays},
	volume = {86},
	copyright = {http://link.aps.org/licenses/aps-default-license},
	issn = {1550-7998, 1550-2368},
	url = {https://link.aps.org/doi/10.1103/PhysRevD.86.102001},
	doi = {10.1103/PhysRevD.86.102001},
	language = {en},
	number = {10},
	urldate = {2025-11-04},
	journal = {Physical Review D},
	author = {Driggers, Jennifer C. and Harms, Jan and Adhikari, Rana X.},
	month = nov,
	year = {2012},
	pages = {102001},
}

@article{harms_optimization_2019,
  author  = {Badaracco, Francesca and Harms, Jan},
  title   = {Optimization of seismometer arrays for the cancellation of {Newtonian} noise from seismic body waves},
  journal = {Class. Quantum Grav.},
  volume  = {36},
  pages   = {145006},
  year    = {2019},
  doi     = {10.1088/1361-6382/ab28c1},
  eprint  = {1903.07936},
  archivePrefix = {arXiv},
  primaryClass  = {astro-ph.IM}
}

@article{badaracco_joint_2024,
	title = {Joint optimization of seismometer arrays for the cancellation of {Newtonian} noise from seismic body waves in the {Einstein} {Telescope}},
	volume = {41},
	issn = {0264-9381, 1361-6382},
	url = {https://iopscience.iop.org/article/10.1088/1361-6382/ad1715},
	doi = {10.1088/1361-6382/ad1715},
	number = {2},
	urldate = {2025-11-04},
	journal = {Classical and Quantum Gravity},
	author = {Badaracco, Francesca and Harms, Jan and Rei, Luca},
	month = jan,
	year = {2024},
	pages = {025013},
}

@article{schillings_fighting_2025,
	title = {Fighting {Newtonian} noise with gradient-based optimization at the {Einstein} {Telescope}},
	volume = {42},
	issn = {0264-9381, 1361-6382},
	url = {https://iopscience.iop.org/article/10.1088/1361-6382/adb898},
	doi = {10.1088/1361-6382/adb898},
	number = {6},
	urldate = {2025-11-04},
	journal = {Classical and Quantum Gravity},
	author = {Schillings, Patrick and Erdmann, Johannes},
	month = mar,
	year = {2025},
	pages = {065025},
}

@article{badaracco_machine_2020,
	title = {Machine learning for gravitational-wave detection: surrogate {Wiener} filtering for the prediction and optimized cancellation of {Newtonian} noise at {Virgo}},
	volume = {37},
	issn = {0264-9381, 1361-6382},
	shorttitle = {Machine learning for gravitational-wave detection},
	url = {https://iopscience.iop.org/article/10.1088/1361-6382/abab64},
	doi = {10.1088/1361-6382/abab64},
	number = {19},
	urldate = {2025-11-04},
	journal = {Classical and Quantum Gravity},
	author = {Badaracco, F and Harms, J and Bertolini, A and Bulik, T and Fiori, I and Idzkowski, B and Kutynia, A and Nikliborc, K and Paoletti, F and Paoli, A and Rei, L and Suchinski, M},
	month = oct,
	year = {2020},
	pages = {195016},
}

@article{allen_detecting_1999,
	title = {Detecting a stochastic background of gravitational radiation: {Signal} processing strategies and sensitivities},
	volume = {59},
	issn = {0556-2821, 1089-4918},
	shorttitle = {Detecting a stochastic background of gravitational radiation},
	url = {http://arxiv.org/abs/gr-qc/9710117},
	doi = {10.1103/PhysRevD.59.102001},
	language = {en},
	number = {10},
	urldate = {2025-08-02},
	journal = {Physical Review D},
	author = {Allen, Bruce and Romano, Joseph D.},
	month = mar,
	year = {1999},
	note = {arXiv:gr-qc/9710117},
	pages = {102001},
}

@article{coughlin_towards_2016,
	title = {Towards a first design of a {Newtonian}-noise cancellation system for {Advanced} {LIGO}},
	volume = {33},
	issn = {0264-9381, 1361-6382},
	url = {https://iopscience.iop.org/article/10.1088/0264-9381/33/24/244001},
	doi = {10.1088/0264-9381/33/24/244001},
	number = {24},
	urldate = {2025-08-02},
	journal = {Classical and Quantum Gravity},
	author = {Coughlin, M and Mukund, N and Harms, J and Driggers, J and Adhikari, R and Mitra, S},
	month = dec,
	year = {2016},
	pages = {244001},
}

@article{flanagan_sensitivity_1993,
	title = {Sensitivity of the {Laser} {Interferometer} {Gravitational} {Wave} {Observatory} to a stochastic background, and its dependence on the detector orientations},
	volume = {48},
	copyright = {http://link.aps.org/licenses/aps-default-license},
	issn = {0556-2821},
	url = {https://link.aps.org/doi/10.1103/PhysRevD.48.2389},
	doi = {10.1103/PhysRevD.48.2389},
	language = {en},
	number = {6},
	urldate = {2025-09-01},
	journal = {Physical Review D},
	author = {Flanagan, Eanna E.},
	month = sep,
	year = {1993},
	pages = {2389--2407},
}

@incollection{casciaro_off-line_2000,
	address = {Milano},
	title = {Off-{Line} {Subtraction} of {Seismic} {Newtonian} {Noise}},
	isbn = {978-88-470-0068-1 978-88-470-2113-6},
	url = {http://link.springer.com/10.1007/978-88-470-2113-6_44},
	language = {en},
	urldate = {2025-09-01},
	booktitle = {Recent {Developments} in {General} {Relativity}},
	publisher = {Springer Milan},
	author = {Cella, G.},
	editor = {Casciaro, B. and Fortunato, D. and Francaviglia, M. and Masiello, A.},
	year = {2000},
	doi = {10.1007/978-88-470-2113-6_44},
	pages = {495--503},
}

@article{coughlin_wiener_2014,
	title = {Wiener filtering with a seismic underground array at the {Sanford} {Underground} {Research} {Facility}},
	volume = {31},
	copyright = {http://iopscience.iop.org/info/page/text-and-data-mining},
	issn = {0264-9381, 1361-6382},
	url = {https://iopscience.iop.org/article/10.1088/0264-9381/31/21/215003},
	doi = {10.1088/0264-9381/31/21/215003},
	number = {21},
	urldate = {2025-09-03},
	journal = {Classical and Quantum Gravity},
	author = {Coughlin, M and Harms, J and Christensen, N and Dergachev, V and DeSalvo, R and Kandhasamy, S and Mandic, V},
	month = nov,
	year = {2014},
	pages = {215003},
}

@article{storn1997differential,
  title={Differential Evolution – A Simple and Efficient Heuristic for Global Optimization over Continuous Spaces},
  author={Storn, Rainer and Price, Kenneth},
  journal={Journal of Global Optimization},
  volume={11},
  number={4},
  pages={341--359},
  year={1997},
  publisher={Springer}
}

@article{hansen2016cma,
  title={The CMA Evolution Strategy: A Tutorial},
  author={Hansen, Nikolaus},
  journal={arXiv preprint arXiv:1604.00772},
  year={2016}
}

@article{virtanen2020scipy,
  title={SciPy 1.0: Fundamental Algorithms for Scientific Computing in Python},
  author={Virtanen, Pauli and Gommers, Ralf and Oliphant, Travis E. and Haberland, Matt and Reddy, Tyler and Cournapeau, David and ... and van der Walt, St{\'e}fan J.},
  journal={Nature Methods},
  volume={17},
  pages={261--272},
  year={2020},
  doi={10.1038/s41592-019-0686-2}
}

@misc{hansen2021pycma,
  title={pycma: CMA-ES for Python},
  author={Hansen, Nikolaus and Auger, Anne},
  year={2021},
  howpublished={\url{https://github.com/CMA-ES/pycma}}
}

@inproceedings{kennedy1995pso,
  author    = {Kennedy, James and Eberhart, Russell},
  title     = {Particle Swarm Optimization},
  booktitle = {Proceedings of the IEEE International Conference on Neural Networks},
  year      = {1995},
  pages     = {1942--1948},
  doi       = {10.1109/ICNN.1995.488968}
}

@article{wales1997basinhopping,
  author    = {Wales, David J. and Doye, Jonathan P. K.},
  title     = {Global Optimization by Basin-Hopping and the Lowest Energy Structures of Lennard-Jones Clusters Containing up to 110 Atoms},
  journal   = {The Journal of Physical Chemistry A},
  volume    = {101},
  number    = {28},
  pages     = {5111--5116},
  year      = {1997},
  doi       = {10.1021/jp970984n}
}

@article{kingma2015adam,
  author    = {Kingma, Diederik P. and Ba, Jimmy},
  title     = {Adam: A Method for Stochastic Optimization},
  journal   = {International Conference on Learning Representations (ICLR)},
  year      = {2015},
  note      = {arXiv:1412.6980},
  url       = {https://arxiv.org/abs/1412.6980}
}

@inproceedings{jose_optimization_2021,
    address = {Iasi, Romania},
    title = {Optimization of {Sensor} {Placement} for {Broadband} {Newtonian} {Noise} {Cancellation} in {GW} {Detectors}},
    copyright = {https://ieeexplore.ieee.org/Xplorehelp/downloads/license-information/IEEE.html},
    isbn = {978-1-6654-1496-8},
    url = {https://ieeexplore.ieee.org/document/9607303/},
    doi = {10.1109/ICSTCC52150.2021.9607303},
    urldate = {2024-12-20},
    booktitle = {2021 25th {International} {Conference} on {System} {Theory}, {Control} and {Computing} ({ICSTCC})},
    publisher = {IEEE},
    author = {Jose, Roselyn and Kalaimani, Rachel Kalpana},
    month = oct,
    year = {2021},
    pages = {132--137},
}

@inproceedings{kasahara_comparison_2018,
	address = {Anaheim, California},
	title = {Comparison of {DAS} (distributed acoustic sensor) and seismometer measurements to evaluate physical quantities in the field},
	url = {https://library.seg.org/doi/10.1190/segam2018-2989627.1},
	doi = {10.1190/segam2018-2989627.1},
	language = {en},
	urldate = {2025-09-08},
	booktitle = {{SEG} {Technical} {Program} {Expanded} {Abstracts} 2018},
	publisher = {Society of Exploration Geophysicists},
	author = {Kasahara, Junzo and Hasada, Yoko and Kawashima, Hirotaka and Sugimoto, Yoshihiro and Yamauchi, Yasutomo and Yamaguchi, Takashi and Yamaguchi, Kenji},
	month = aug,
	year = {2018},
	pages = {191--195},
}

@misc{rading_distributed_2025,
	title = {Distributed {Acoustic} {Sensing} for {Environmental} {Monitoring}, and {Newtonian} {Noise} {Mitigation}:{Comparable} {Sensitivity} to {Seismometers}},
	copyright = {Creative Commons Attribution 4.0 International},
	shorttitle = {Distributed {Acoustic} {Sensing} for {Environmental} {Monitoring}, and {Newtonian} {Noise} {Mitigation}},
	url = {https://arxiv.org/abs/2507.13523},
	doi = {10.48550/ARXIV.2507.13523},
	urldate = {2025-09-08},
	publisher = {arXiv},
	author = {Rading, Reinhardt and Badaracco, Fracensca and Beis, Spiridon and Isleif, Katharina Sophie and Ophardt, Paul and Vossius, Wanda and Collaboration, the WAVE},
	year = {2025},
	note = {Version Number: 1},
}

@article{shinohara_performance_2022,
	title = {Performance of {Seismic} {Observation} by {Distributed} {Acoustic} {Sensing} {Technology} {Using} a {Seafloor} {Cable} {Off} {Sanriku}, {Japan}},
	volume = {9},
	issn = {2296-7745},
	url = {https://www.frontiersin.org/articles/10.3389/fmars.2022.844506/full},
	doi = {10.3389/fmars.2022.844506},
	urldate = {2025-09-08},
	journal = {Frontiers in Marine Science},
	author = {Shinohara, Masanao and Yamada, Tomoaki and Akuhara, Takeshi and Mochizuki, Kimihiro and Sakai, Shin’ichi},
	month = apr,
	year = {2022},
	pages = {844506},
}

@article{rossi_assessment_2022,
	title = {Assessment of {Distributed} {Acoustic} {Sensing} ({DAS}) performance for geotechnical applications},
	volume = {306},
	issn = {00137952},
	url = {https://linkinghub.elsevier.com/retrieve/pii/S0013795222002149},
	doi = {10.1016/j.enggeo.2022.106729},
	language = {en},
	urldate = {2025-09-08},
	journal = {Engineering Geology},
	author = {Rossi, Matteo and Wisén, Roger and Vignoli, Giulio and Coni, Mauro},
	month = sep,
	year = {2022},
	pages = {106729},
}

@article{paitz_empirical_2021,
	title = {Empirical {Investigations} of the {Instrument} {Response} for {Distributed} {Acoustic} {Sensing} ({DAS}) across 17 {Octaves}},
	volume = {111},
	issn = {0037-1106, 1943-3573},
	url = {https://pubs.geoscienceworld.org/ssa/bssa/article/111/1/1/592023/Empirical-Investigations-of-the-Instrument},
	doi = {10.1785/0120200185},
	language = {en},
	number = {1},
	urldate = {2025-09-08},
	journal = {Bulletin of the Seismological Society of America},
	author = {Paitz, Patrick and Edme, Pascal and Gräff, Dominik and Walter, Fabian and Doetsch, Joseph and Chalari, Athena and Schmelzbach, Cédric and Fichtner, Andreas},
	month = feb,
	year = {2021},
	pages = {1--10},
}

@book{vaseghi_advanced_2000,
	address = {Chichester},
	edition = {Second ed},
	title = {Advanced {Digital} {Signal} {Processing} and {Noise} {Reduction} ({Second} {Edition})},
	isbn = {978-0-470-84162-4},
	language = {eng},
	publisher = {John Wiley Sons Ltd},
	author = {Vaseghi, Saeed V.},
	collaborator = {{John Wiley \& Sons, Inc}},
	year = {2000},
}

\appendix
\section{Wiener Filter for a fusion sensor array}
\label{appendix:fusionWF}

% ---------- Appendix: Wiener Filter for a fusion sensor array ----------
%\section*{Wiener Filter for a fusion sensor array}
In this appendix we show how to calculate the Wiener filter for a fusion sensor array.
Using the correlation coefficients defined in table \ref{tab:correlation_coefficients_explicit}, we can write down the cross-spectrum matrix as well as the cross-correlation vector.
For the sensors, we use four seismometers and two strainmeters placed as shown in table \ref{tab:fusion_array_positions}. % and Fig.~\ref{fig:...}.

% Positions (from 'flat'):
\begin{table}[h!]
  \centering
  \caption{Sensor positions (units as in the main text).}
  \label{tab:fusion_array_positions}
  \begin{tabular}{c c r r r}
    \hline
    ID & Type & $x$ & $y$ & $z$ \\
    \hline
    1 & Seismometer & 0.0000 & \phantom{-}0.28355112 & 0.0000 \\
    2 & Seismometer & 0.0000 & -0.28355112          & 0.0000 \\
    3 & Seismometer & 0.0000 & \phantom{-}0.00000000 & 0.28355112 \\
    4 & Seismometer & 0.0000 & \phantom{-}0.00000000 & -0.28355112 \\
    5 & Strainmeter & 0.10347379 & 0.00000000 & 0.00000000 \\
    6 & Strainmeter & -0.10347379 & 0.00000000 & 0.00000000 \\
    \hline
  \end{tabular}
\end{table}

% Cross-spectrum matrix with visible block boundaries (no extra packages needed)
We denote the sensor cross-spectrum matrix by $\mathbf{C}_{SS}$ and the cross-correlation vector by $\mathbf{C}_{SN}$.
With the ordering (1–4)~seismometers and (5–6)~strainmeters, the block structure is
\[
\mathbf{C}_{SS} \;=\;
\begin{bmatrix}
\mathbf{C}_{SS}^{\mathrm{\boldsymbol{\xi}}} & \mathbf{C}_{SS}^{\mathrm{\boldsymbol{\xi},\boldsymbol{\epsilon}}} \\
\big(\mathbf{C}_{SS}^{\mathrm{\boldsymbol{\xi},\boldsymbol{\epsilon}}}\big)^{\!\top} & \mathbf{C}_{SS}^{\mathrm{\boldsymbol{\epsilon}}}
\end{bmatrix}
\;=\;
\left[
\begin{array}{rrrr|rr}
 0.3348 & -0.0221 & -0.0122 & -0.0122 & \phantom{-}0.0422 & -0.0422 \\
 -0.0221 & \phantom{-}0.3348 & -0.0122 & -0.0122 & \phantom{-}0.0422 & -0.0422 \\
 -0.0122 & -0.0122 & \phantom{-}0.3348 & -0.0221 & \phantom{-}0.0422 & -0.0422 \\
 -0.0122 & -0.0122 & -0.0221 & \phantom{-}0.3348 & \phantom{-}0.0422 & -0.0422 \\
 \hline
 \phantom{-}0.0422 & \phantom{-}0.0422 & \phantom{-}0.0422 & \phantom{-}0.0422 & \phantom{-}0.1116 & \phantom{-}0.0450 \\
 -0.0422 & -0.0422 & -0.0422 & -0.0422 & \phantom{-}0.0450 & \phantom{-}0.1116
\end{array}
\right].
\]

The cross-correlation vector (sensor–to–target) is
\[
\mathbf{C}_{SN} =
\begin{bmatrix}
 0.0504 \\ 0.0504 \\ 0.0504 \\ 0.0504 \\ 0.0140 \\ -0.0140
\end{bmatrix}.
\]

\section{Algorithmic Hyperparameters and Runtime Benchmarks}\label{appendix: CMA-ES eval}

The DE + CMA-ES optimization runs were performed on a MacBook Pro equipped with an Apple M4 Pro CPU. All benchmarks reported below were obtained using single-threaded runs, in order to 
ensure reproducibility.

\begin{table}[h!]
    \centering
    \caption{Average runtime of the DE + CMA-ES optimization algorithm for different numbers of seismometers $N$. Each value corresponds to the mean wall-clock time of ten independent runs.}
    \label{tab:runtime}
    \begin{tabular}{c c}
        \hline
        Number of seismometers $N$ & Runtime [s] \\
        \hline
        6  & 6.172 \\
        7  & 6.816 \\
        8  & 9.654 \\
        9  & 10.266 \\
        10 & 13.985 \\
        \hline
    \end{tabular}
\end{table}

The Differential Evolution (implemented in Scipy \cite{virtanen2020scipy}) stage was run with a population size of 10 per dimension ($\text{pop} = \text{popsize} \times D$, $D = 3$ as we have 3 spatial dimensions), using the \texttt{rand1bin} strategy, recombination constant of 0.7, and Latin hypercube initialization. The optimization was terminated with a tolerance of $10^{-6}$, without polishing, and monitored via an early-switch callback with $100$ iterations of warm-up, a patience parameter of 25, and an improvement tolerance of $10^{-5}$. The DE solution space is restricted to $\mathcal{B} = [l_b, u_b]^3$, the bounds typically needs to be found by fine-tuning. The Covariance Matrix Adaptation Evolution Strategy (CMA-ES) (implemented by using the package \cite{hansen2021pycma}) stage was initialized with the best candidate from DE (or its final result if no switch occurred), and with an initial step size $\sigma_0 = 0.07 \langle u_b - l_b \rangle$ based on the average parameter range. CMA-ES was configured with population size $4 + 3 \log D$, bounds set by the parameter limits $[l_b, u_b]$, and stopping criterion $\texttt{tolfun} = 10^{-5}$, with a maximum of 200 iterations.

\end{document}